\documentclass[final,12pt]{elsarticle_Z}

\usepackage{graphicx}
\usepackage{amssymb}
\usepackage{amsthm}
\usepackage{lineno}

\usepackage{subfig,bm,amsmath,array,url}
\renewcommand{\b}{\mathbf}

\newcommand{\R}{\mathbb{R}}

\newtheorem*{lemma}{Lemma}

\journal{Digital Signal Processing}

\begin{document}
\begin{frontmatter}

\title{Distributed High Dimensional Information Theoretical Image Registration via Random Projections\tnoteref{t1}}
\tnotetext[t1]{\copyright{} 2012 Elsevier Inc.\ Digital Signal Processing 22(6):894-902, 2012. The original publication is available at \url{http://dx.doi.org/10.1016/j.dsp.2012.04.018}.}
\author{Zolt\'{a}n Szab\'{o}}\ead{szzoli@cs.elte.hu}\corref{cor1}
\author{Andr\'{a}s L\H{o}rincz}\ead{andras.lorincz@elte.hu}
\cortext[cor1]{Corresponding author. Fax: +36 1 381 2140.}

\address{E\"{o}tv\"{o}s Lor{\'a}nd University, Department of Software Technology and Methodology P\'{a}zm\'{a}ny P{\'e}ter s{\'e}t{\'a}ny 1/C, Budapest, H-1117, Hungary}

\begin{abstract}%100words
Information theoretical measures, such as entropy, mutual information, and various divergences,
exhibit robust characteristics in image registration applications. However, the estimation of these
quantities is computationally intensive in high dimensions. On the other hand, consistent
estimation from pairwise distances of the sample points is possible, which suits random projection
(RP) based low dimensional embeddings. We adapt the RP technique to this task by means of a simple
ensemble method. To the best of our knowledge, this is the first distributed, RP based information
theoretical image registration approach. The efficiency of the method is demonstrated through
numerical examples.
\end{abstract}


\end{frontmatter}
%\linenumbers

\section{Introduction}
Machine learning methods are notoriously limited by the high dimensional nature of the data. This
problem may be alleviated via the random projection (RP) technique, which has been successfully
applied, e.g., in the fields of classification
\cite{fradkin03experiments,deegalla06reducting,goel05face}, clustering \cite{fern03random},
independent subspace analysis \cite{szabo09fast}, search for approximate nearest neighbors
\cite{ailon06approximate}, dimension estimation of manifolds \cite{hegde07random}, estimation of
geodesic paths \cite{mahmoudi08estimation}, learning mixture of Gaussian models
\cite{dasgupta00experiments}, compression of image and text data \cite{bingham01random}, data
stream computation \cite{li07nonlinear,krishna07incremental} and reservoir computing
\cite{likosevicius09reservoir}. For a recent RP review, see \cite{vempala04random}. We note that
the RP technique is closely related to the signal processing method of compressed sensing
\cite{baraniuk08simple}.

As it has been shown recently in a number of works
\cite{ozuysal09fast,kokiopoulou09optimal,balin09accelerating,healy07fast}, the RP approach has
potentials in patch classification and image registration. For example, \cite{ozuysal09fast}
combines the votes of random binary feature groups (ferns) for the classification of random patches
in a naive Bayes framework. Promising registration methods using $L_1$ and $L_2$ (Euclidean
distance, correlation) norms have been introduced in \cite{kokiopoulou09optimal} and
\cite{balin09accelerating,healy07fast}, respectively.

Information theoretical cost functions, however, exhibit more robust properties in multi-modal
image registration \cite{neemuchwala07image,kybic04high,bardera06high}. Papers
\cite{neemuchwala07image,kybic04high} apply k-nearest neighbor based estimation. However, the
computation of these quantities is costly in high dimensions \cite{arya98optimal} and the different
image properties (e.g., colors, intensities of neighborhood pixels, gradient information, 
output of spatial filters, texture descriptors) may easily lead to
high dimensional representation. The task is formulated as the estimation of \emph{discrete} mutual
information in \cite{bardera06high} and the solution is accomplished by equidistant sampling of $D$
points from randomly positioned straight lines. The method estimates a histogram of $N^{2D}$ bins,
where $N$ is the number of bins of the image, which may considerably limit computational
efficiency.

Here we address the problem of information theoretical image registration in case of high
dimensional features. Particularly, we demonstrate that Shannon's multidimensional differential
entropy can be efficiently estimated for high dimensional image registration purposes through RP
methods. Our solution enables distributed evaluation. The presented approach extends the method
presented in \cite{szabo09fast} in the context of independent subspace analysis (ISA)
\cite{cardoso98multidimensional}, where we exploited the fact that ISA can be formulated as the
optimization problem of the sum of entropies under certain conditions
\cite{szabo07undercomplete_TCC}. Here, to our best knowledge, we present the first distributed RP
based information theoretical image registration approach.

The paper is structured as follows: In Section~\ref{sec:background} we shortly review the image
registration problem as well as the method of random projections. Section~\ref{sec:method}
formulates our RP based solution idea for image registration. Section~\ref{sec:illustrations}
contains the numerical illustrations. Conclusions are drawn in Section~\ref{sec:conclusions}.

\section{Background}\label{sec:background}
First, we describe the image registration task (Section~\ref{sec:imreg}) followed by low distortion
embeddings and random projections (Section~\ref{sec:RP}).

\subsection{The Image Registration Problem}\label{sec:imreg}
In image registration one has two images, $\b{I}_{\text{ref}}$ and $\b{I}_{\text{test}}$, as
well as a family of geometrical transformations, such as scaling, translation, affine
transformations, and warping. We assume that the transformations can be described by some parameter
$\bm{\theta}$ and let  $\bm{\Theta}$ denote the set of the possible parameters.  Let transformation
with parameter $\bm{\theta}\in \bm{\Theta}$ on $\b{I}_{\text{test}}$ produce
$\b{I}_{\text{test}}(\bm{\theta})$. The goal of image registration is to find the transformation
(parameter $\bm{\theta}$) for which the warped test image $\b{I}_{\text{test}}(\bm{\theta})$ is the
`closest' possible to reference image $\b{I}_{\text{ref}}$. Formally, the task is
\begin{equation*}
J(\bm{\theta})=S(\b{I}_{\text{ref}},\b{I}_{\text{test}}(\bm{\theta}))\rightarrow \max_{\bm{\theta}\in\bm{\Theta}},
\end{equation*}
where the similarity of two images is given by the similarity measure $S$. Registration depends on the
similarity measure and one use -- among other things -- $L_2$ and $L_1$ norm, or different
information theoretical similarity measures.

Let feature $f(p;\b{I})\in\R^D$ denote the feature of image $\b{I}$ associated with pixel
$p\in{\b{I}}$. In the simplest case, the feature is the pixel itself, but one
can choose a neighborhood of the pixel, edge information at and around the pixel, the RGB values
for colored images, or combinations of these. For registrations based on the $L_q$ norm $(q\in\{1,2\})$, the cost
function takes the form
\begin{equation*}
J_q(\bm{\theta})=-\sum_{p}\left\|f(p;\b{I}_{\text{ref}})-f(p;\b{I}_{\text{test}}(\bm{\theta}))\right\|_q,
\end{equation*}
where for a vector $\b{v}\in\R^D$  $\left\|\b{v}\right\|_q=(\sum_{i=1}^D|v_i|^q)^{1/q}$. Instead of the similarity of features in $\left\|\cdot\right\|_1$ and
$\left\|\cdot\right\|_2$ norms, one might consider similarity by means of information theoretical
concepts. An example is that we take the negative value of the joint entropy of the features of
images $\b{I}_{\text{ref}}$, $\b{I}_{\text{test}}(\bm{\theta})$ as our cost function
\cite{garcia08regional}:
\begin{equation}
J_H(\bm{\theta}) = -H(\b{I}_{\text{ref}},\b{I}_{\text{test}}(\bm{\theta})),\label{eq:cost}
\end{equation}
where $H$ denotes Shannon's multidimensional differential entropy \cite{cover91elements}. One may
replace entropy $H$ in \eqref{eq:cost} by other quantities, e.g., by the R\'enyi's $\alpha$-entropy,
the $\alpha$-mutual information, and the $\alpha$-divergence, to mention some of the candidate similarity
measures \cite{neemuchwala07image}.

\subsection{Low Distortion Embeddings, Random Projection}\label{sec:RP}
Low distortion embedding and random projections are relevant for our purposes. Low distortion
embedding intends to map a set of points of a high dimensional Euclidean space to a much lower
dimensional one by preserving the distances between the points approximately. Such low dimensional
approximate isometric embedding exists according to the Johnson-Lindenstrauss Lemma
\cite{johnson84extensions}:

\begin{lemma}[Johnson-Lindenstrauss]  Given a number $\varepsilon\in (0,1)$ and a point set $\{\b{v}_1,\ldots,\b{v}_T\}\subset\R^D$
of $T$ elements. Then for $d=O(\ln(T)/\varepsilon^2)$ there exists a \mbox{Lipschitz} mapping $f:\R^D\rightarrow\R^d$ such that
\begin{equation}
(1-\varepsilon)\left\|\b{v}_i-\b{v}_j\right\|_2 \le \left\|f(\b{v}_i)-f(\b{v}_j)\right\|_2\le (1+\varepsilon)\left\|\b{v}_i-\b{v}_j\right\|_2 \label{eq:JL}
\end{equation}
for any $1\le i,j\le T$.
\end{lemma}
During the years, a number of explicit constructions have appeared for the construction of $f$.
Notably, one can show that the property embraced by \eqref{eq:JL} is satisfied with probability that approaches 1 for random
linear mapping ($f(\b{v})=\b{Pv}$, $\b{P}\in\R^{d\times D}$) provided that $\b{P}$ is chosen to
project to a random $d$-dimensional subspace \cite{frankl98johnson}.

Less strict conditions on $\b{P}$ are also sufficient and many of them decreases computational
costs. Introducing the notation $\b{P}=\frac{1}{\sqrt{d}}\b{R}$, i.e.,
$f(\b{v})=\frac{1}{\sqrt{d}}\b{Rv}$, $\b{R}\in\R^{d\times D}$, it is sufficient that matrix
elements $r_{ij}$ of matrix $\b{R}=[r_{ij}]$ are drawn independently from the $N(0,1)$
\emph{standard normal} distribution \cite{indyk98approximate}.\footnote{Multiplier
$\frac{1}{\sqrt{d}}$ in expression $\b{P}=\frac{1}{\sqrt{d}}\b{R}$ means that the length of the
rows of matrix $\b{P}$ is not strictly one; it is sufficient if their lengths are 1 on the
average.} Other explicit constructions for $\b{R}$ include Rademacher and  (very) sparse distributions 
\cite{achlioptas03database,li06very}. More general methods are also available based on weak moment 
constraints \cite{arriga06algorithmic,matousek08variants}.

\section{Method}\label{sec:method}
In image registration information theoretical registration measures show robust
characteristics when compared with $L_1$ and $L_2$ measures\footnote{The family of $L_2$ measures
include the correlation defined by the scalar product.}, e.g., in
\cite{kybic04high} directly for \eqref{eq:cost}, and for $\alpha$-entropy, $\alpha$-mutual
information, and $\alpha$-divergence in \cite{neemuchwala07image}. However, these estimations have
high computational burdens since the dimension of the features in the cited references are $25$ and
$64$ respectively. Here, we deal with the efficient estimation of cost function \eqref{eq:cost}
that from now on we denote by $J$. We note that the idea of efficient estimation can be used for a
number of information theoretical quantities, provided that they can be estimated by means of
pairwise Euclidean distances of the samples.

Central to our RP based distributed method are the following:
\begin{enumerate}
    \item
    The computational load can be decreased by
    \begin{enumerate}
      \item dividing the samples into groups and then
      \item computing the averages of the group estimates \cite{kybic04high}.
    \end{enumerate}
    We call this the \emph{ensemble approach}.
    \item Estimation of the multidimensional entropy cost function $J$ can be carried out consistently by
    \emph{nearest neighbor methods} using pairwise Euclidean distances of sample points
    \cite{kozachenko87statistical,hero02application,leonenko08class}.
\end{enumerate}
Taking into account that low dimensional approximate isometric embedding of points of high
dimensional Euclidean space can be addressed by the Johnson-Lindenstrauss Lemma and the related
random projection methods, we suggest the following procedure for distributed RP based entropy (and
thus $J$) estimation:
\begin{enumerate}
    \item divide the $T$ feature samples\footnote{In the image registration task the set of feature samples is
    $\{[f(p,\b{I}_{\text{ref}});f(p,\b{I}_{\text{test}}(\bm{\theta}))]\}$ where $p$ is the running index and $[\b{a};\b{b}]$ denotes
    the concatenation of vectors $\b{a}$ and $\b{b}$.} $\{\b{v}(1),\ldots,\b{v}(T)\}\subset\R^D$ into $N$ groups indexed
    by sets $I_1,\ldots,I_N$ so that each group contains $G$ samples,
    \item for all fixed groups take the random projection of $\b{v}$ as
        \begin{equation*}
            \b{v}_{n}^{\text{RP}}(t):=\b{R}_n \b{v}(t)\quad  (t\in I_n; \,\, n=1,\ldots,N; \,\, \b{R}_n\in\R^{d\times D}), \label{eq:v_n}
        \end{equation*}
    Note: normalization factor $\frac{1}{\sqrt{d}}$ can be dropped in $\b{P}=\frac{1}{\sqrt{d}}\b{R}$
    since it becomes an additive constant term for the case of the differential entropy, $H(c\b{v})=H(\b{v})+d\log(|c|)$.
    \item average the estimated entropies of the RP-ed groups to get the estimation
    \begin{equation}
        \hat{H}(\b{v})=\frac{1}{N}\sum_{n=1}^N\hat{H}(\b{v}_{n}^{\text{RP}}).\label{eq:H-hat}
    \end{equation}
\end{enumerate}

In the next section we illustrate the efficiency of the proposed RP based approach in 
image registration.

\section{Illustrations}\label{sec:illustrations}
\begin{figure}%[h!]
\centering%
\subfloat[][]{\includegraphics[width=10cm]{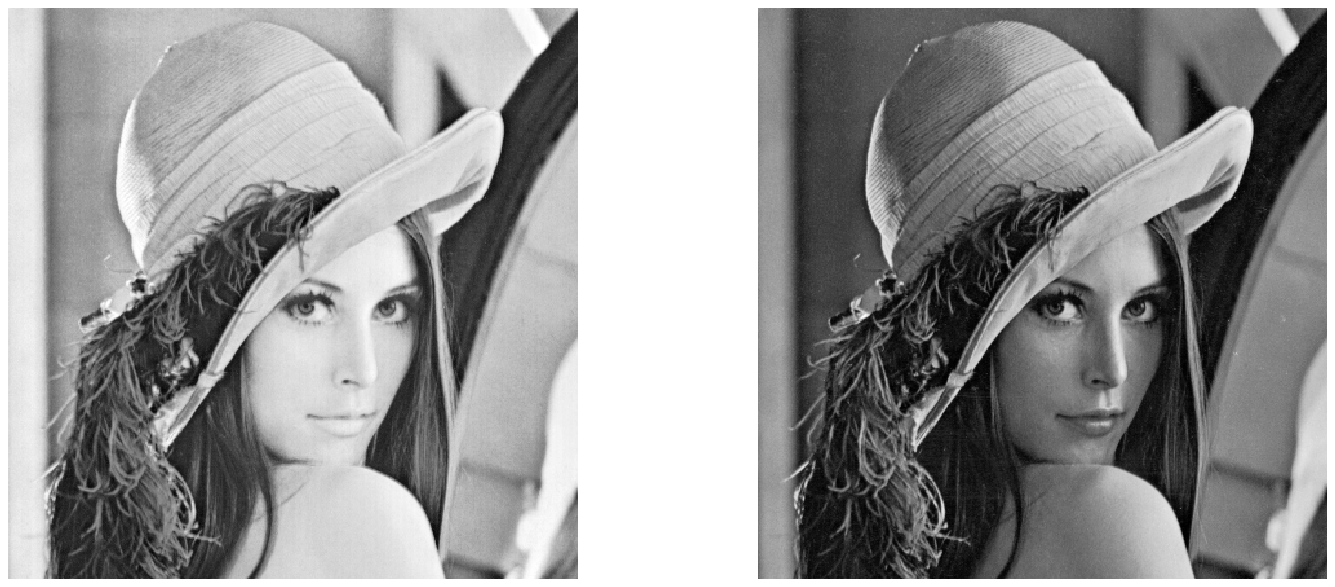}}\\%
\subfloat[][]{\includegraphics[width=10cm]{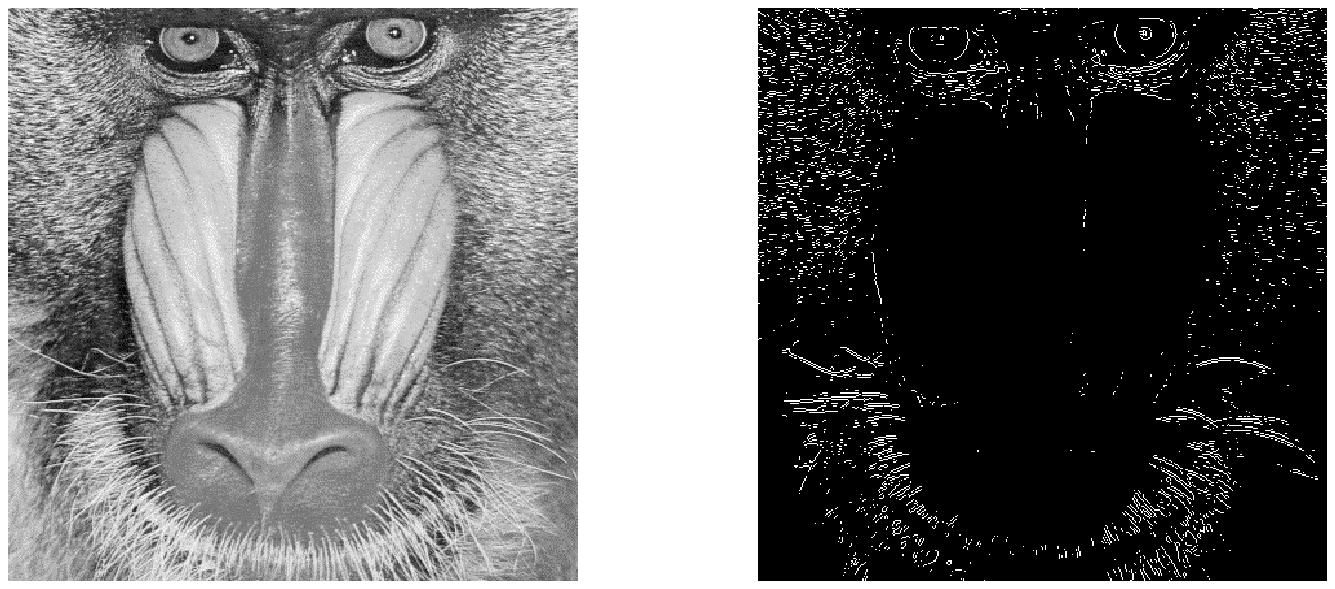}}%
\caption[Illustration of the \emph{Lena} and \emph{mandrill} databases.]
{Illustration of the (a): \emph{Lena} test, (b): \emph{mandrill} test.}%
\label{fig:databases}%
\end{figure}

In our illustrations, we show examples that enable quantitative evaluation and reproduction:
\begin{enumerate}
 \item We use $512\times 512$ images:
    \begin{enumerate}
     \item
      In test \emph{Lena}, we register the rotated versions of the red and green
      channels of image \emph{Lena}, see Fig.~\ref{fig:databases}(a).
     \item
      In the \emph{mandrill} test we register the rotated versions of the gray-scale image of a \emph{mandrill} baboon and
      its Sobel filtered version, see Fig.~\ref{fig:databases}(b).
    \end{enumerate}
 \item we chose to evaluate the objective function \eqref{eq:cost} for angles from $-10^{\circ}$ to $10^{\circ}$ by steps
 $0.5^{\circ}$ and in interval $[-1^{\circ},1^{\circ}]$ by steps $0.1^{\circ}$. In the ideal case the optimal degree
 $\theta^{*}$ is $0$. Our performance measure is the deviation from the optimal value.
\end{enumerate}

In our simulations,
\begin{itemize}
 \item
    we chose the $D=(2h+1)\times (2h+1)$ rectangle around each pixel as the feature $f$ of that pixel.
 \item
    coordinates of the $\b{R}_n$ RP matrices were drawn independently from standard normal distribution, but more
    general constructions could also be used \cite{arriga06algorithmic,matousek08variants}.
 \item
    for each individual parameter, $10$ random runs were averaged. Our parameters included
    $h$, the linear size of the neighborhood that determines dimension $D$ of the  feature, $G$, the
    size of the randomly projected groups and $d$, the dimension of RP.
  \item
    performance statistics are summarized by means of notched boxed plots, which show the quartiles ($Q_1,Q_2,Q_3$), depict the
    outliers, i.e., those that fall outside of interval $[Q_1 - 1.5(Q_3-Q_1), Q_3 + 1.5(Q_3-Q_1)]$ by
    circles, and whiskers represent the largest and smallest non-outlier data points.
  \item
    we studied the efficiency of five different entropy estimating methods in \eqref{eq:H-hat} including
    \begin{itemize}
      \item the recursive k-d partitioning scheme \cite{stowell09fast}, 
      \item the $k$-nearest neighbor method \cite{leonenko08class},
      \item generalized k-nearest neighbor graphs \cite{david10estimation},
      \item minimum spanning tree based entropy graphs \cite{yukich98probability,hero02application} and
      \item the weighted nearest neighbor method \cite{sricharan11weighted}. 
    \end{itemize}
    The methods will be referred to as \emph{kdp}, \emph{kNN$_k$}, \emph{kNN$_{1-k}$}, \emph{MST} and 
    \emph{wkNN}, respectively. 
    \emph{kdp} is a plug-in type method estimating the underlying density directly, hence especially efficient for small dimensional 
      ($d$) problems. \cite{david10estimation} extends the approach of \cite{leonenko08class} ($S=\{k\}$) to an arbitrary 
      $S$ neighborhood subset ($S\subseteq\{1,\ldots,k\}$). In our experiments, we set $S=\{1,\ldots,k\}$. Instead of k-nearest graphs the 
      total sum of pairwise distances is minimized over spanning trees in the MST method. 
      The \emph{kNN$_k$}, \emph{kNN$_{1-k}$}, \emph{MST} constructions belong to the 
      general umbrella of quasi-additive functionals \cite{yukich98probability} providing statistically 
      consistent estimation for the R{\'e}nyi  entropy ($H_{\alpha}$) \cite{renyi61measures} .
     The Shannon's entropy $H$ is a special case of this family since
     $\lim_{\alpha\rightarrow 1}H_{\alpha}=H$. In our simulations, we chose $\alpha=0.95$. $k$, the number of neighbors in kNN$_k$, kNN$_{1-k}$ was $5$.
     Finally, the \emph{wkNN} technique makes use of a weighted combination of k-nearest neighbor estimators for different $k$ values.
  \item 
      $h$, the neighborhood parameter was selected from the set $\{1,2,3,4,5,10,20,30\}$.
  \item
    $G$, the size of groups and $d$, the RP dimension  took values $20$, $50$, $100$, $1,000$, $10,000$, $100,000$ and $1,2,5,8$, respectively.
  \item
     the $T$ feature points were distributed randomly into groups of size $G$ in order to increase the diversity of
     the individual groups.
\end{itemize}

\begin{figure}%[h!]
  \centering%
  \subfloat[][]{\includegraphics[width=7.15cm]{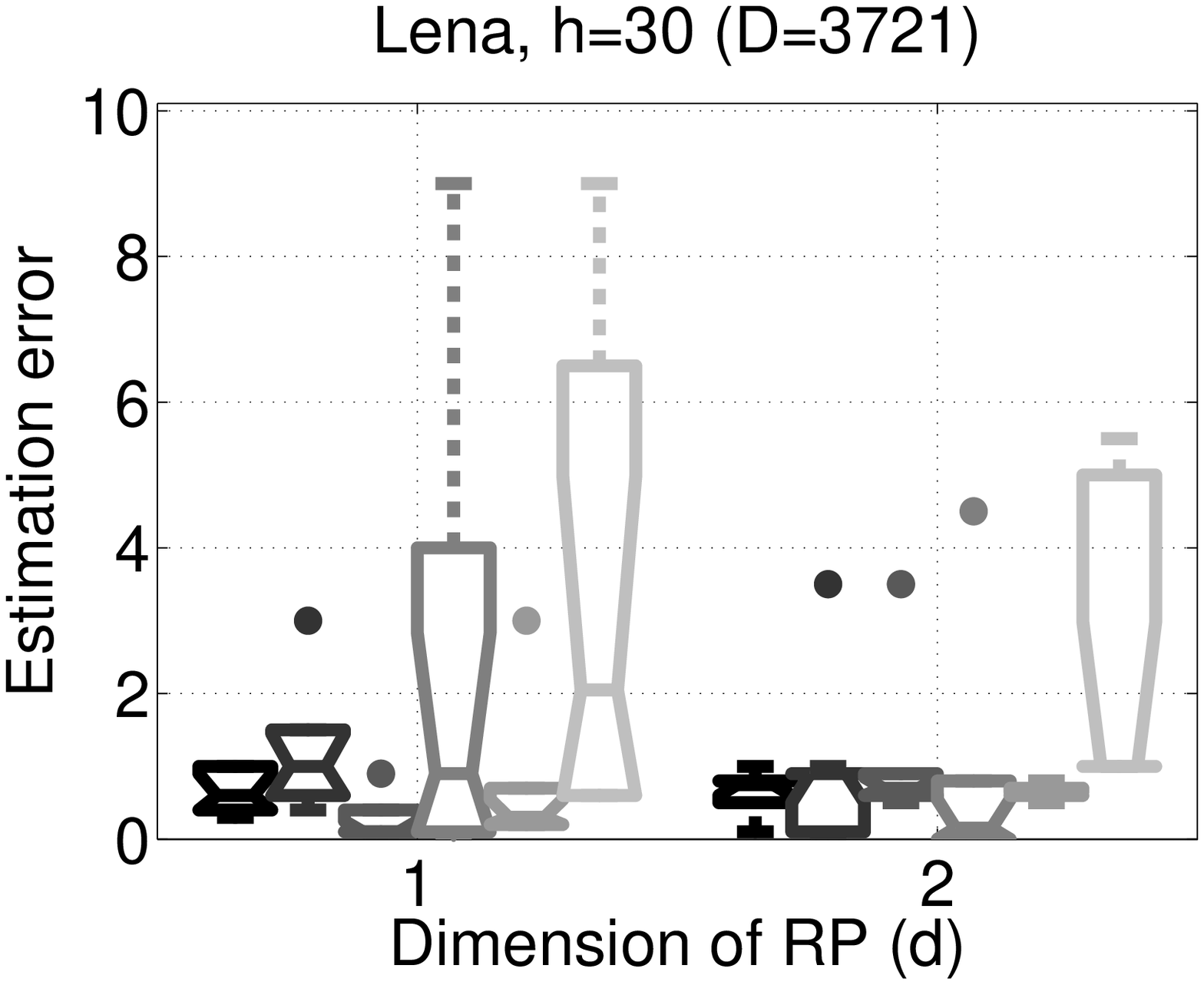}}%
  \subfloat[][]{\includegraphics[width=7.15cm]{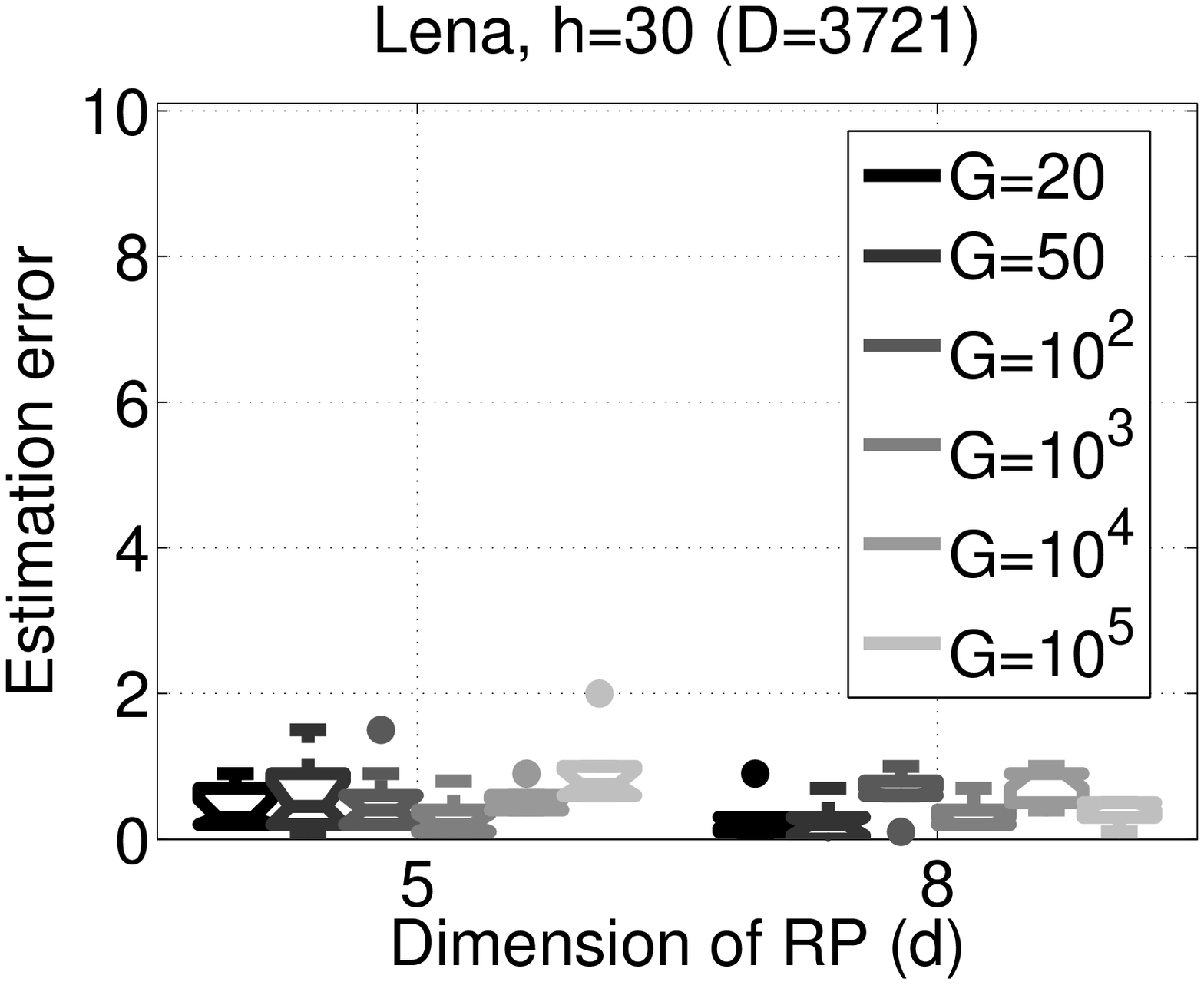}}\\%
  \subfloat[][]{\includegraphics[width=7.15cm]{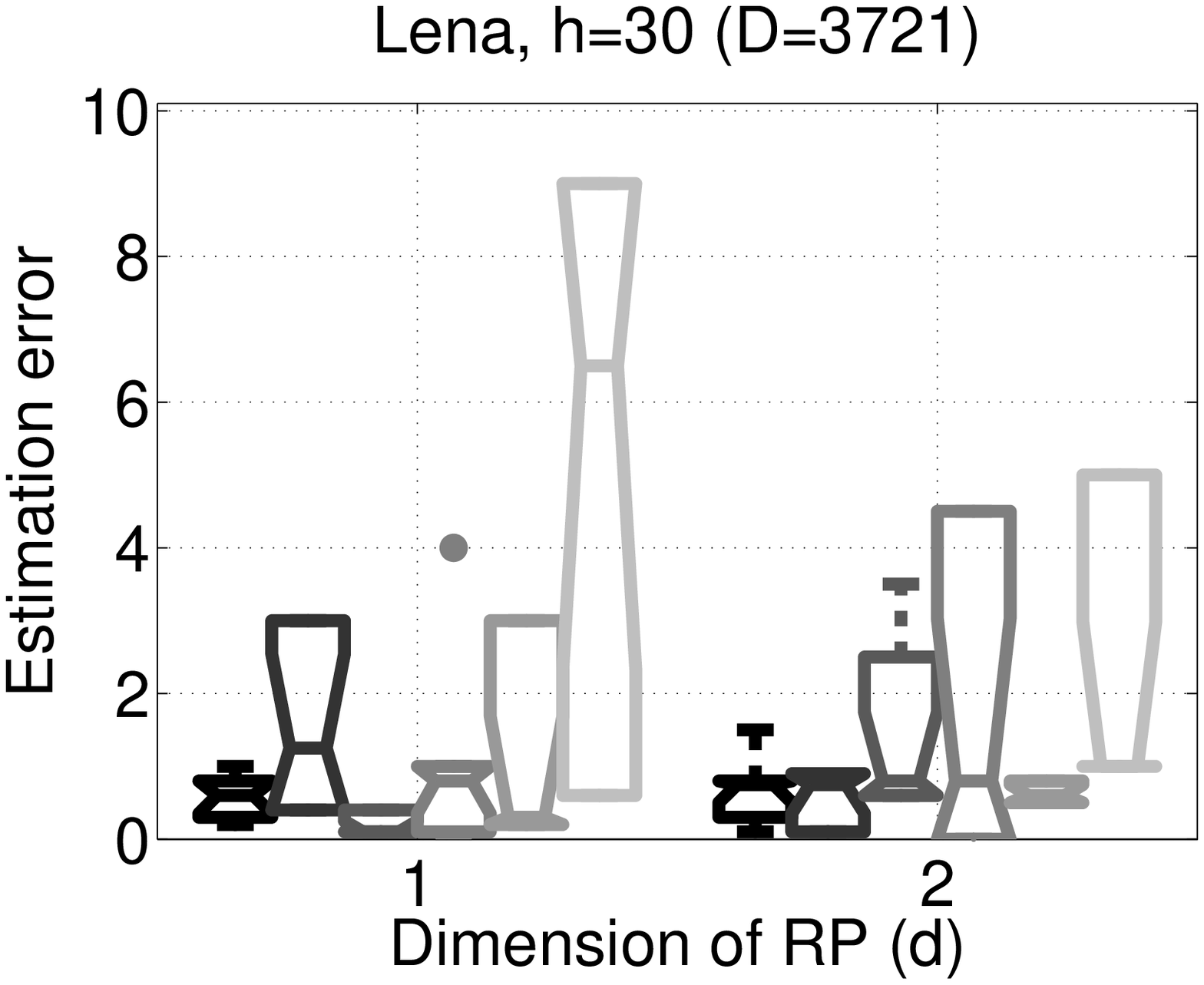}}%
  \subfloat[][]{\includegraphics[width=7.15cm]{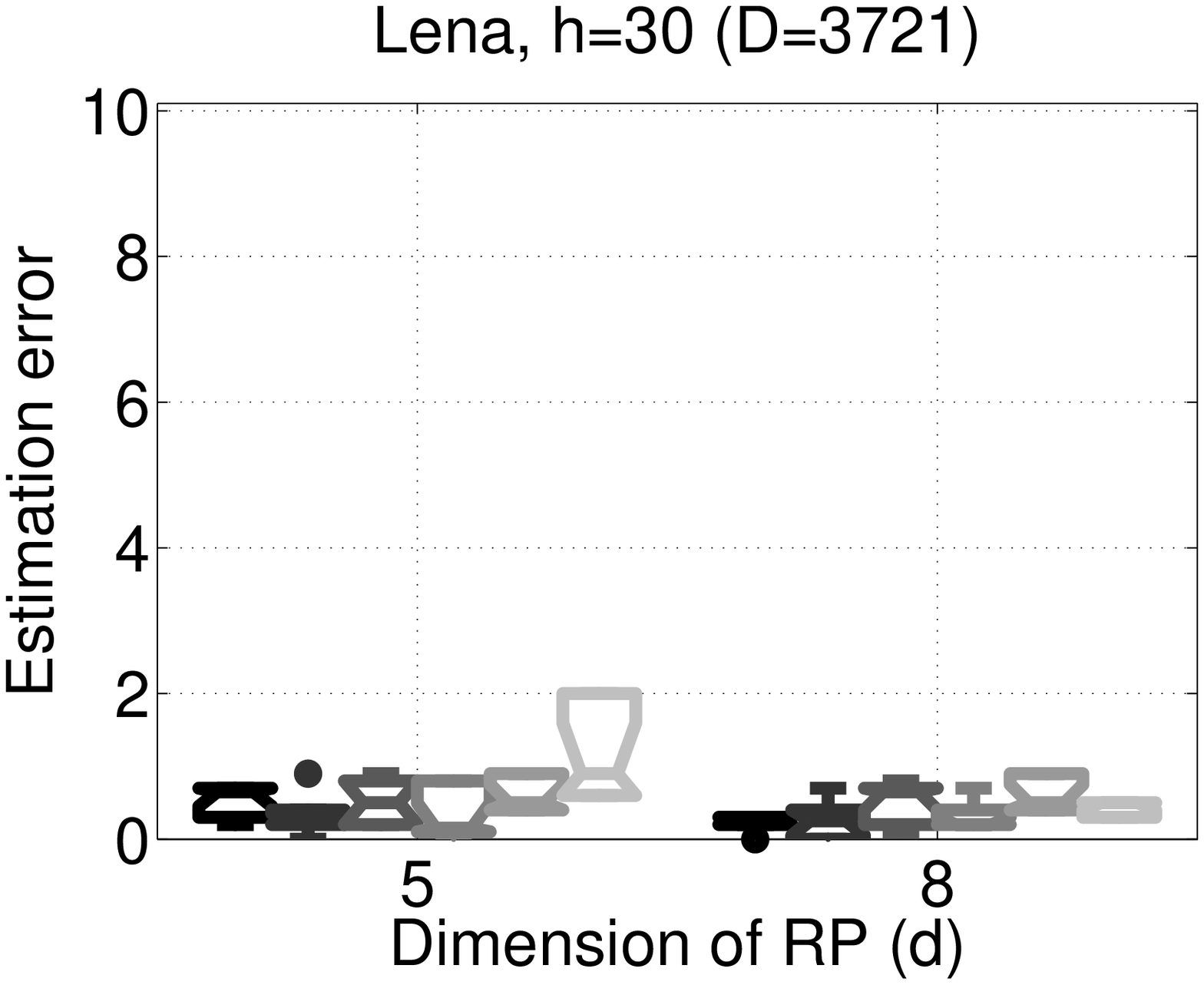}}%
  \caption[Estimation error: \emph{Lena} dataset, \emph{kNN$_{k}$} and \emph{kNN$_{1-k}$} methods.]{
    Estimation error as a function of the RP dimension $d$ on the \emph{Lena} dataset for different $G$ group sizes. 
    Neighbor size: $h=30$. (a)-(b): \emph{kNN$_k$} method. (c)-(d): \emph{kNN$_{1-k}$} method. First column: $d=1$, $2$. Second column: $d=5$, $8$.}%
  \label{fig:lena:h30:kNNk-kNN1tok}%
\end{figure}

\begin{figure}%[h!]
  \centering%
  \subfloat[][]{\includegraphics[width=7.15cm]{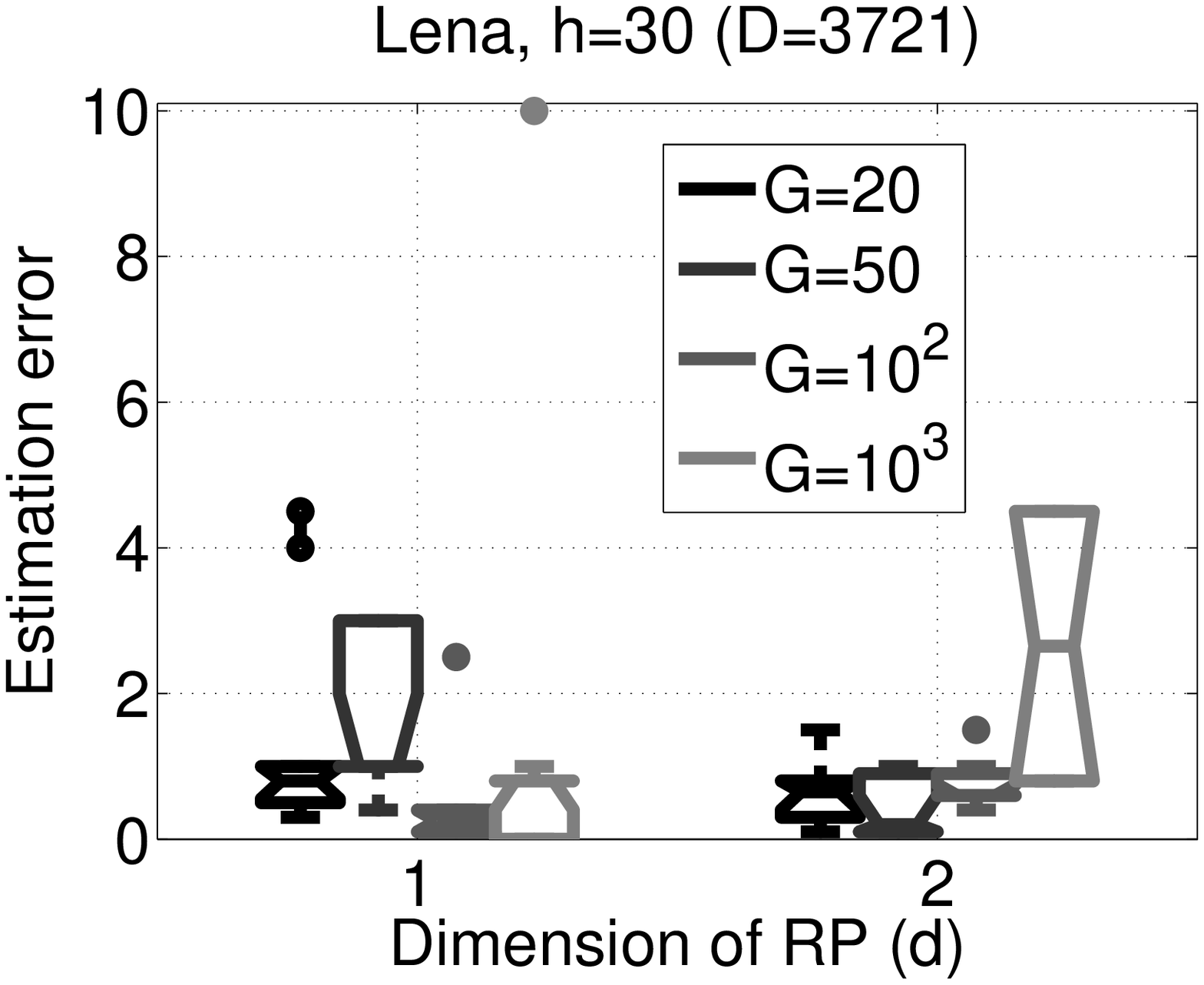}}%
  \subfloat[][]{\includegraphics[width=7.15cm]{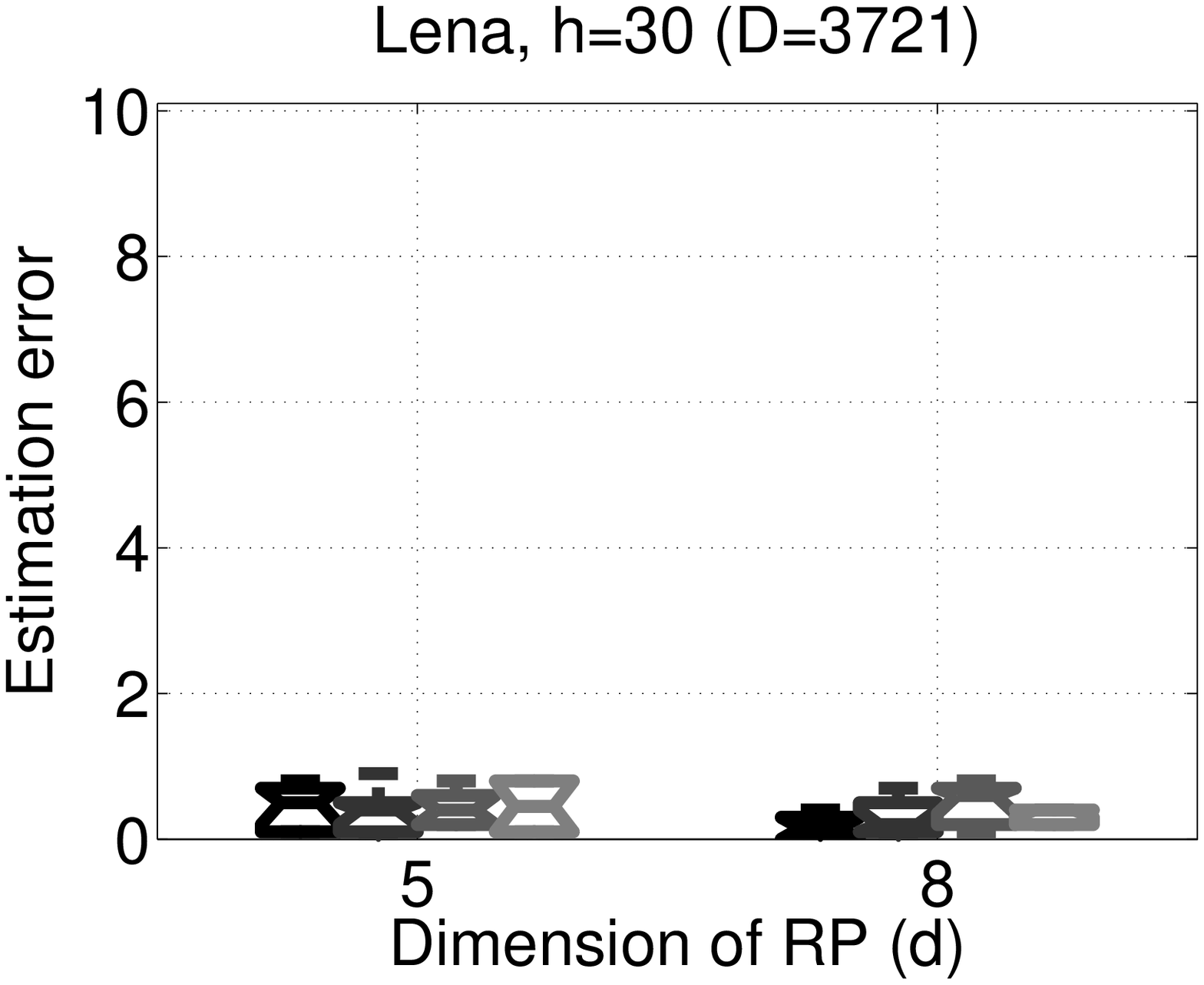}}\\%magic_const=2
  \subfloat[][]{\includegraphics[width=7.15cm]{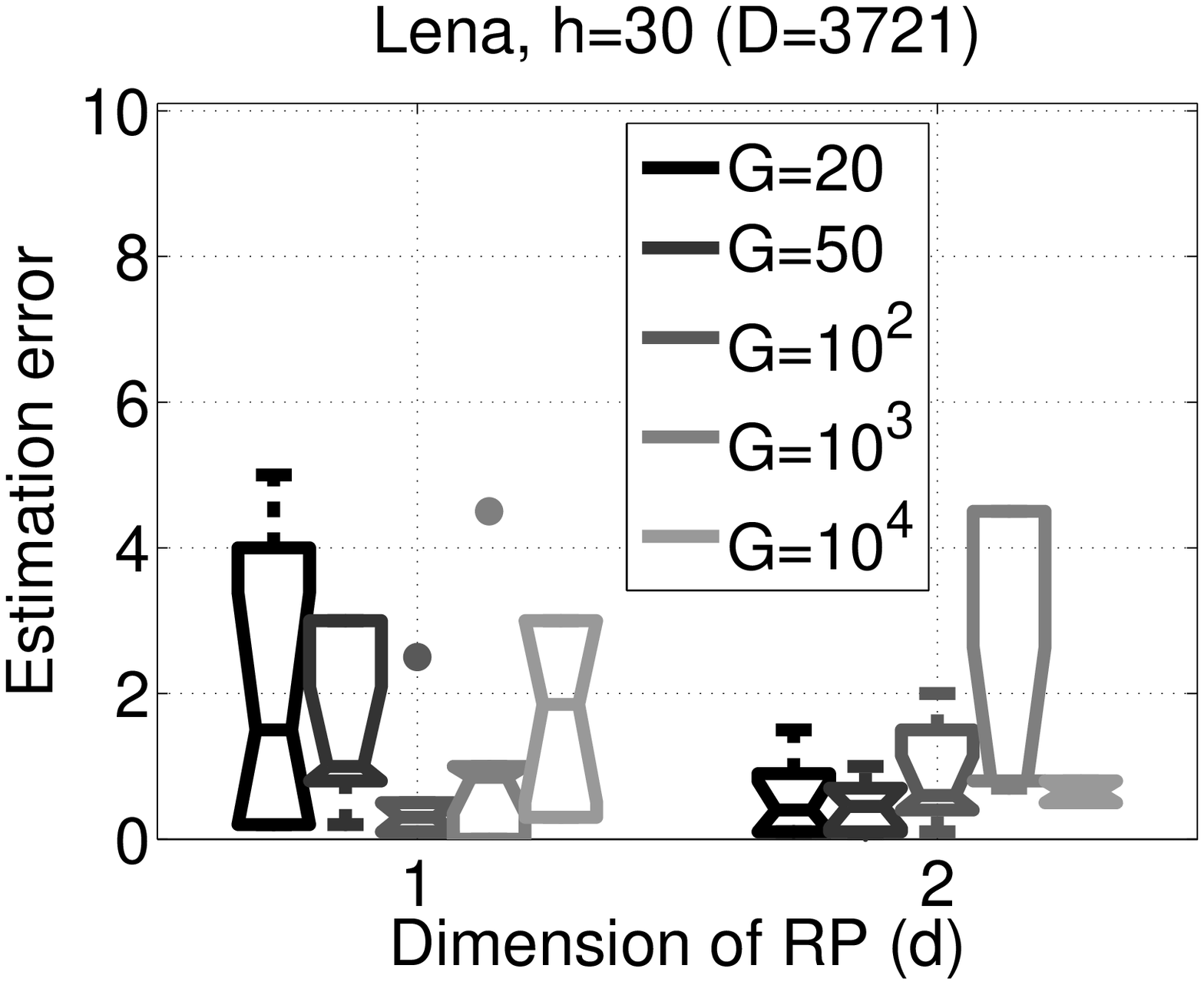}}%
  \subfloat[][]{\includegraphics[width=7.15cm]{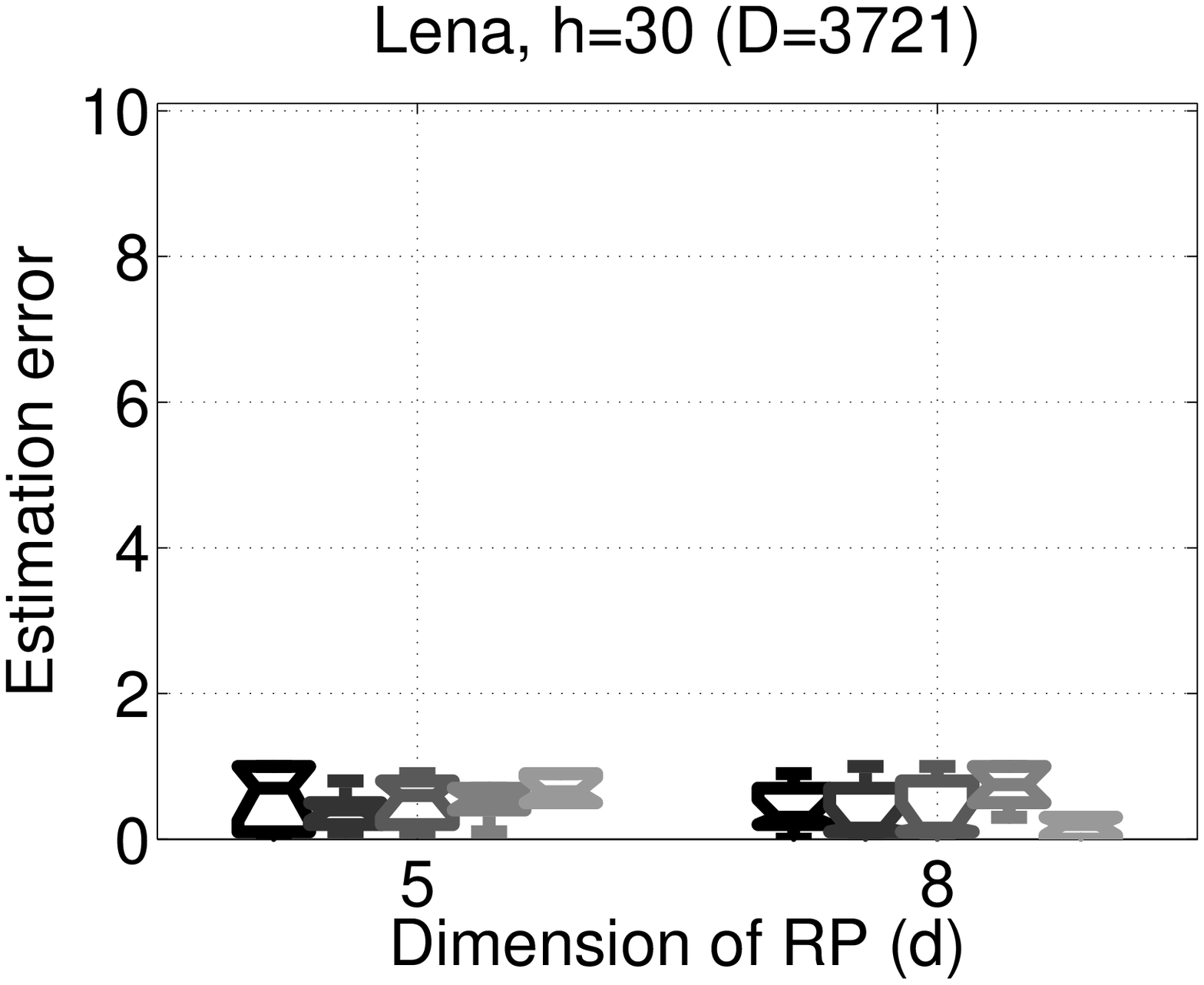}}%%magic_const=2.3
  \caption[Estimation error: \emph{Lena} dataset, \emph{MST} and \emph{wkNN} methods.]{
    Estimation error as a function of the RP dimension $d$ on the \emph{Lena} dataset for different $G$ group sizes. 
    Neighbor size: $h=30$. (a)-(b): \emph{MST} method. (c)-(d): \emph{wkNN} method. First column: $d=1$, $2$. Second column: $d=5$, $8$.}%
  \label{fig:lena:h30:weightedkNN-MST}%
\end{figure}

In the \textbf{first set of experiments} we focused on the precision of the estimations on the \emph{Lena} dataset.
According to our experiences 
\begin{itemize}
 \item there is no relevant/visible difference in the precision of the estimations for $h\in\{1,2,3,4,5,10\}$. The estimation is even of high precision for
  $h=10$ that we illustrate in Fig.~\ref{fig:lena:h10,30:kdpee}(a)-(b) for the \emph{kdp} technique. The estimation errors are quite similar 
  for  $h=20$ and $30$, the latter is shown in Fig.~\ref{fig:lena:h10,30:kdpee}(c)-(d). Here, one can
  notice a small uncertainty in the estimations for smaller RP dimensions ($d=1,2$), which is moderately 
  present   for larger $d$ values ($d=5,8$) -- except for the largest studied group size $G=100,000$.
 \item the obtained results are of similar precision on this dataset for all the studied entropy estimators. 
    We illustrate this property for the most challenging $h=30$ value in 
  Fig.~\ref{fig:lena:h30:kNNk-kNN1tok} and Fig.~\ref{fig:lena:h30:weightedkNN-MST}.
\end{itemize}

\begin{figure}%[h!]
  \centering%
  \subfloat[][]{\includegraphics[width=7.15cm]{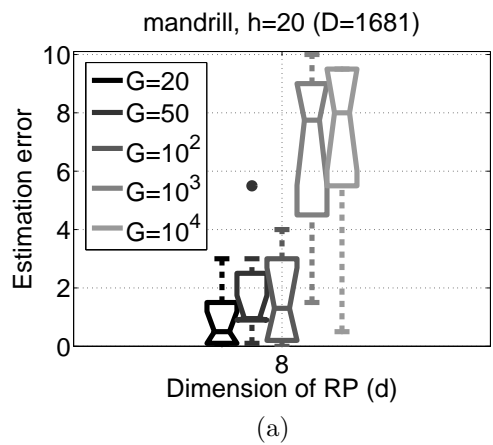}}%
  \caption[Estimation error: \emph{mandrill} dataset, \emph{wkNN} method ($h=20$).]{Estimation error for RP dimension $d=8$ and for different $G$ group sizes. Method: \emph{wkNN}. 
    Neighbor size: $h=20$.}%
  \label{fig:mandrill:h20:weightedkNN}%
\end{figure}

In the \textbf{second set of experiments} we were dealing with the \emph{mandrill} dataset, where different
modalities of the same image (pixel, edge filtered version) had to be registered. Here,
\begin{itemize}
 \item 
  the \emph{kdp} approach gradually deteriorates as the dimension of the underlying feature 
  representation is increasing, i.e., as a function of $h$. For $h=1$, the method
  gives precise estimations for $d=1,2$ and small group sizes ($G=20,50,100$); other parameter 
  choices result in uncertain estimations, see Fig.~\ref{fig:mandrill:h1,5:kdpee}(a)-(b). By increasing the size of 
  neighborhood ($h$), the estimations \emph{gradually} break down. For $h=5$, the precisions are depicted in Fig.~\ref{fig:mandrill:h1,5:kdpee}(c)-(d); the estimations are still acceptable.
  For $h=10$ we did not obtain valuable estimations for the \emph{kdp} technique.
\item in contrast to the \emph{kdp} method, the \emph{kNN$_{k}$}, \emph{kNN$_{1-k}$}, \emph{MST} 
    and \emph{wkNN} techniques are all capable of coping with the $h=5$ and $h=10$ values, as it is illustrated 
    in Fig.~\ref{fig:mandrill:h5,10:kNNk}, Fig.~\ref{fig:mandrill:h5,10:kNN1tok}, 
    Fig.~\ref{fig:mandrill:h5,10:MST} and Fig.~\ref{fig:mandrill:h5,10:weightedkNN}, respectively. 
    It can also be observed, that the RP dimension must be $d\ge 2$ here, and in case of $d=5,8$ one obtaines
    highly precise/certain estimations.
\item the only method which could cope with the increased $h=20$ neighbor size value, was the 
    $wkNN$ technique. This result could be achieved for RP dimension $d=8$ making use of small group sizes 
    ($G=20, 50, 100$), see Fig.~\ref{fig:mandrill:h20:weightedkNN}.
\end{itemize}

\begin{figure}%[h!]
  \centering%
  \subfloat[][]{\includegraphics[width=9.5cm]{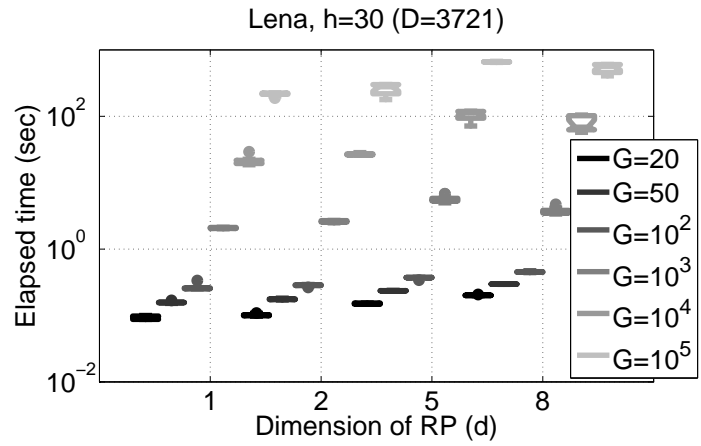}}\\%
  \subfloat[][]{\includegraphics[width=9.5cm]{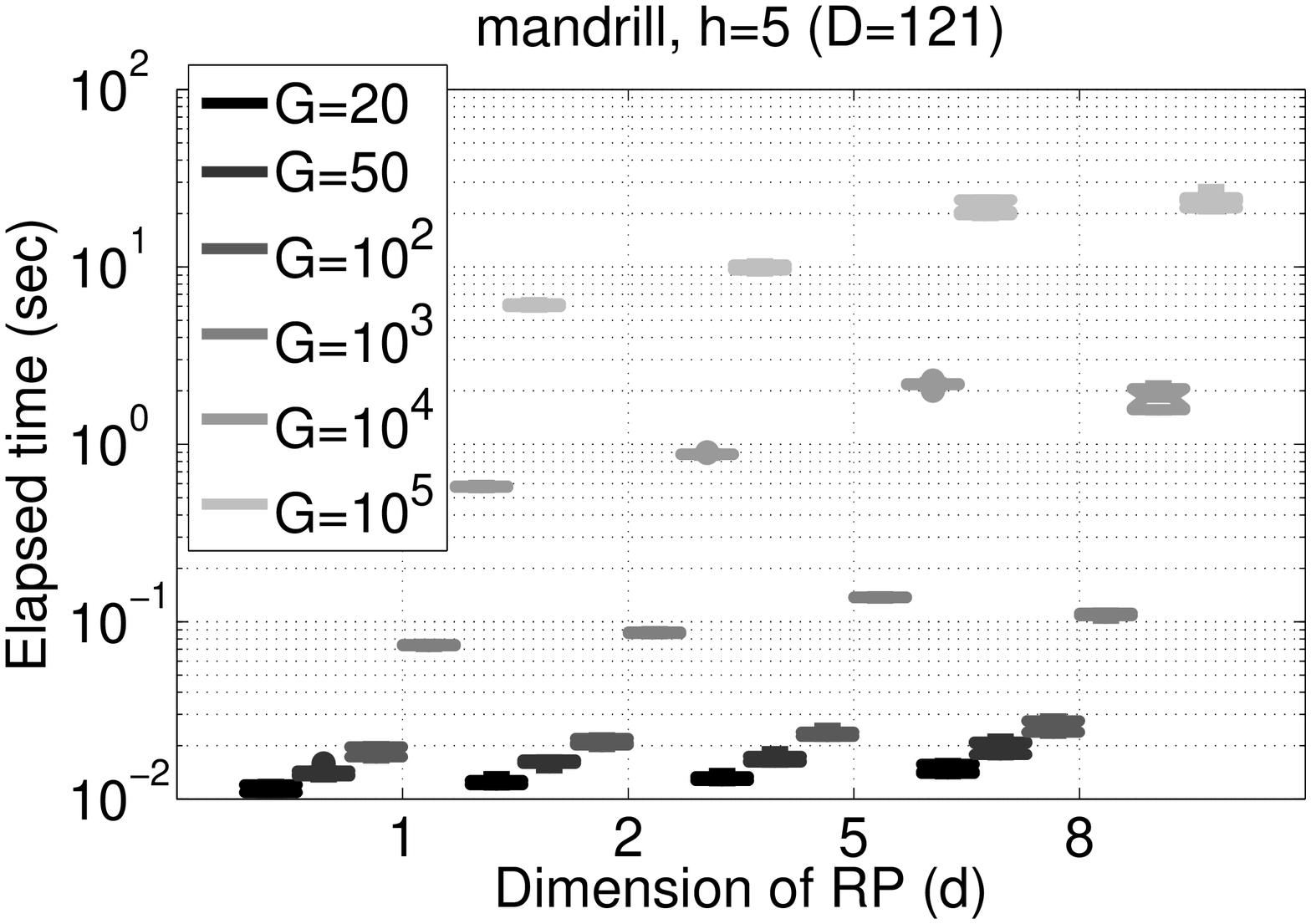}}%
  \caption[Elapsed time: \emph{Lena} ($h=30$), \emph{mandrill} ($h=5$)]{Computation time 
      as a function of the RP dimension for different $G$ group sizes with log scale on the $y$ axis. Method: \emph{kdp}. (a): \emph{Lena} dataset, neighbor size $h=30$. (b): \emph{mandrill} dataset, $h=5$.}%
  \label{fig:kdpee:elapsed}%
\end{figure}

The \textbf{computation times} are illustrated for the \emph{Lena} ($h=30$) and \emph{mandrill} dataset ($h=5$) for 
the \emph{kdp} method in Fig.\ref{fig:kdpee:elapsed}(a) and Fig.\ref{fig:kdpee:elapsed}(b), respectively.
As it can be seen, the ensemble approach with group size $G=20-100$ may speed up computations by
several orders of magnitudes; similar trends can be obtained for the other estimators, too.
Among the studied methods, the \emph{kdp} technique was the most competitive in terms of computation time.
We also present the computation times for the largest studied problem, \emph{Lena} with $h=30$; compared to \emph{kdp}
\begin{itemize}
  \item the \emph{kNN$_k$} and \emph{kNN$_{1-k}$} techniques were within a factor of $1.5-2.5$ in terms of computation time,
  \item the \emph{wkNN} method was $1.5-2$ ($3-4$) times slower compared to the \emph{kdp} approach in case of $G\le 1,000$ ($G=10,000$), and
  \item the \emph{MST} based estimator was within a factor of $1.5-2$ compared to \emph{kdp} in case of $G\le 100$, and more than $6$ times slower for $G=1,000$.
\end{itemize}
As it can be seen in Fig.~\ref{fig:kdpee:elapsed}, the application of the reduced RP dimension can be advantageous in terms of computation time.
Moreover, compared to schemes without dimensionality reduction ($d=D$ and $\b{R}_n=\b{I}$, $\forall n$), i.e., working directly on raw data,
the presented RP based dimensionality approach can heavily speed-up computations. This behaviour is already 
present for $h=5$, as it is illustrated for $d=2$ on the \emph{mandrill} dataset in Table~\ref{tab:elapsed-vs-d2}.

Considering the possible $\mathbf{(h,d,G)}$ \textbf{choices}, according to our numerical experiences,
\begin{itemize}
 \item often, small $d=2-5$ RP dimensions give rise to reliable estimations for several entropy methods, 
 \item it is necessary to slowly increase $d$ as a function of the dimension of the feature representation 
    (parameterized by $h$),
 \item in the studied parameter domain, group sizes of $G=50-100$ could provide precise estimations, and 
    simultaneously open the door to massive speed-up by distributed solutions.
\end{itemize}

These results demonstrate the efficiency of our RP based approach.

\section{Conclusions}\label{sec:conclusions}
We have shown that the random projection (RP) technique can be adapted to distributed information
theoretical image registration. Our extensive numerical experiments including five different entropy 
estimators demonstrated that the proposed approach (i) can offer orders of magnitude in computation time, and (ii) provides
robust estimation for large dimensional features.

It is very promising since it is parallel and fits multi-core architectures,
including graphical processors. Since information theoretical measures are robust, our method may be
useful in diverse signal processing areas with the advance of multi-core hardware.

\section*{Acknowledgments}
The European Union and the European Social Fund have provided financial
support to the project under the grant agreement no.\ T{\'A}MOP
4.2.1./B-09/1/KMR-2010-0003. The research has also been supported by
the `European Robotic Surgery' EC FP7 grant (no.: 288233). Any
opinions, findings and conclusions or recommendations expressed in this
material are those of the authors and do not necessarily reflect the
views of other members of the consortium or the European Commission.

The authors would like to thank to Kumar Sricharan for making available
the implementation of the wkNN method.

\bibliographystyle{model1-num-names}
\bibliography{main}

\vspace*{1cm}

%short biography:
\noindent\textbf{About the author}--ZOLT\'{A}N SZAB\'{O} (Applied Mathematics M.Sc. 2006, Ph.D. 2012,
Informatics Ph.D. 2009) is a research fellow at the E\"{o}tv\"{o}s Lor{\'a}nd University. In 2007, he won the Scientist of the Year 
Award of the Faculty of Informatics. In 2008, he obtained the Bronze Medal of the Pro 
Patria et Scientia Award of Hungarian Ph.D. Students. He is a reviewer at the IEEE Transactions 
on Neural Networks and Learning Systems, Signal, Image and Video Processing, Neurocomputing and 
IEEE Transactions on Signal Processing journals. His research interest include Independent Subspace Analysis and its extensions, 
information theory, kernel methods, group-structured dictionary learning and collaborative filtering.\\\\
\textbf{About the author}--ANDR\'{A}S L\H{O}RINCZ (Physics M.Sc. 1975, Solid
State Physics Ph.D. 1978, Molecular Physics C.Sc. 1986, Laser Physics
habilitation, 1998, Information Technology habilitation, 2009) is a senior
researcher of Information Science at E{\"o}tv{\"o}s Lor{\'a}nd University.
He is a Fellow of the European Coordinating Committee for Artificial
Intelligence. He has published more than 140 peer reviewed journal 80 peer
reviewed conference  papers on his research areas. He has been leading a
group working on different aspects of intelligent systems.

%\newpage
%\listoffigures

%\newpage

%-----------------
%estimation error: lena:

\begin{figure}%[h!]
  \centering%
  \subfloat[][]{\includegraphics[width=7.15cm]{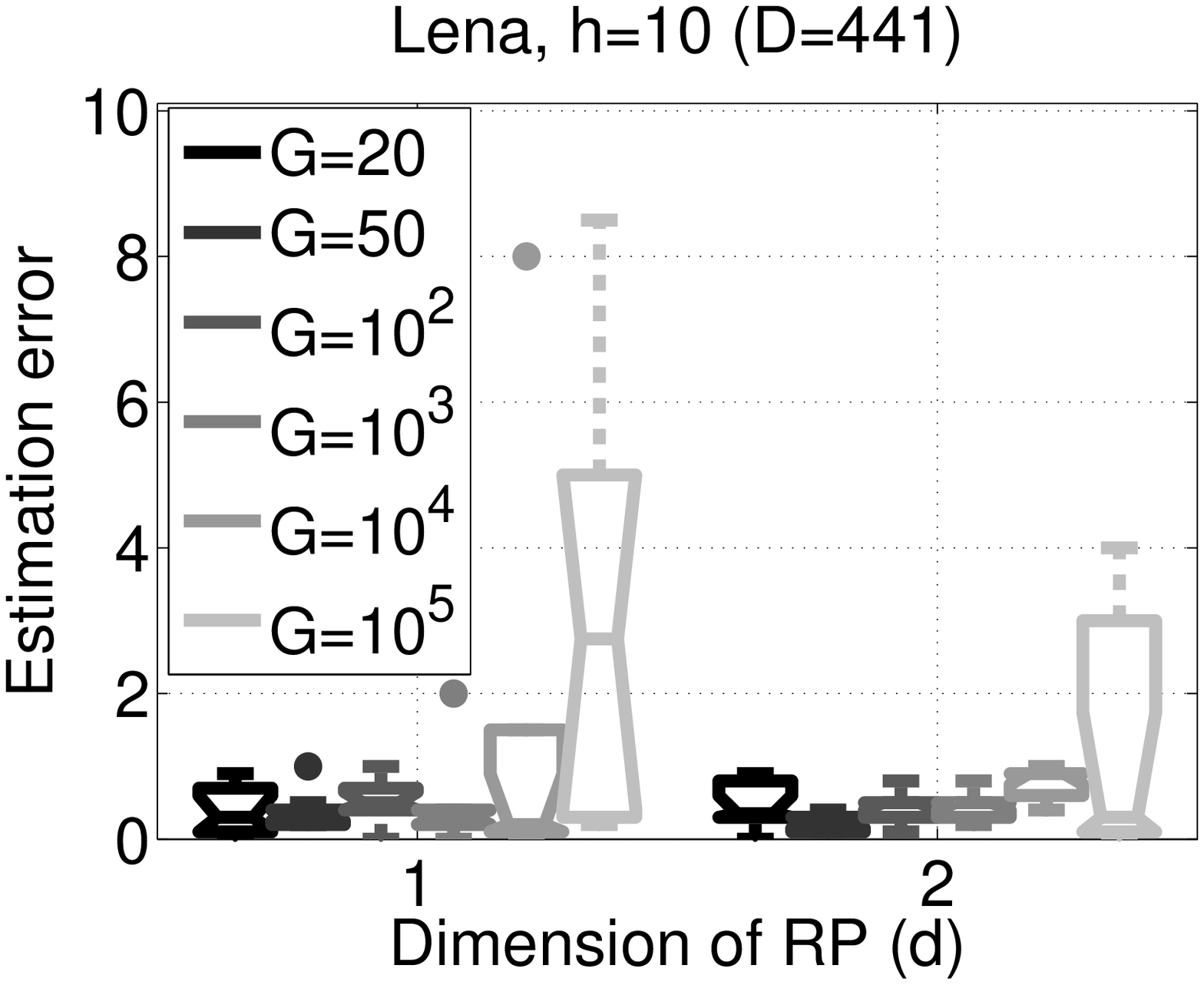}}%
  \subfloat[][]{\includegraphics[width=7.15cm]{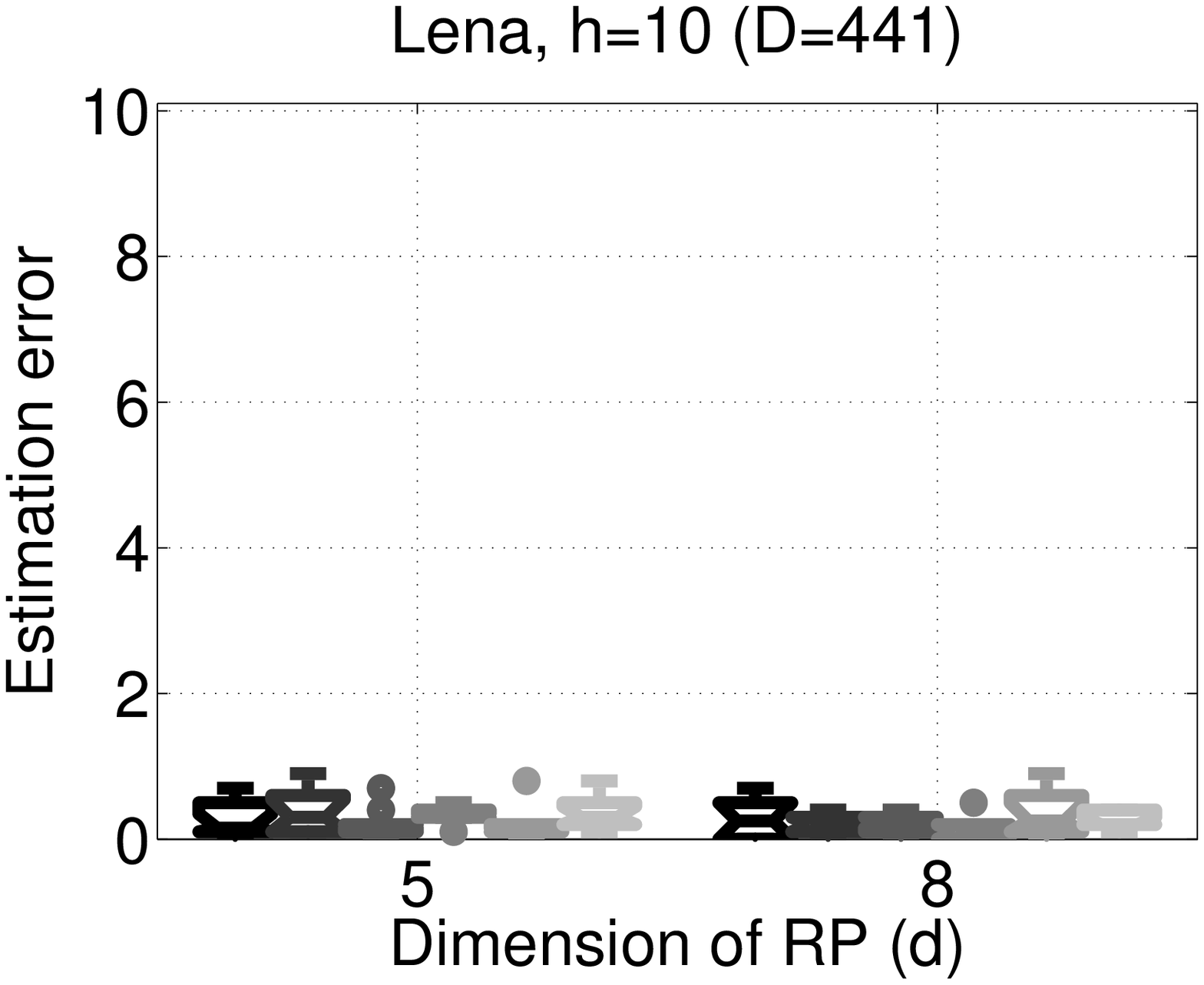}}\\%
  \subfloat[][]{\includegraphics[width=7.15cm]{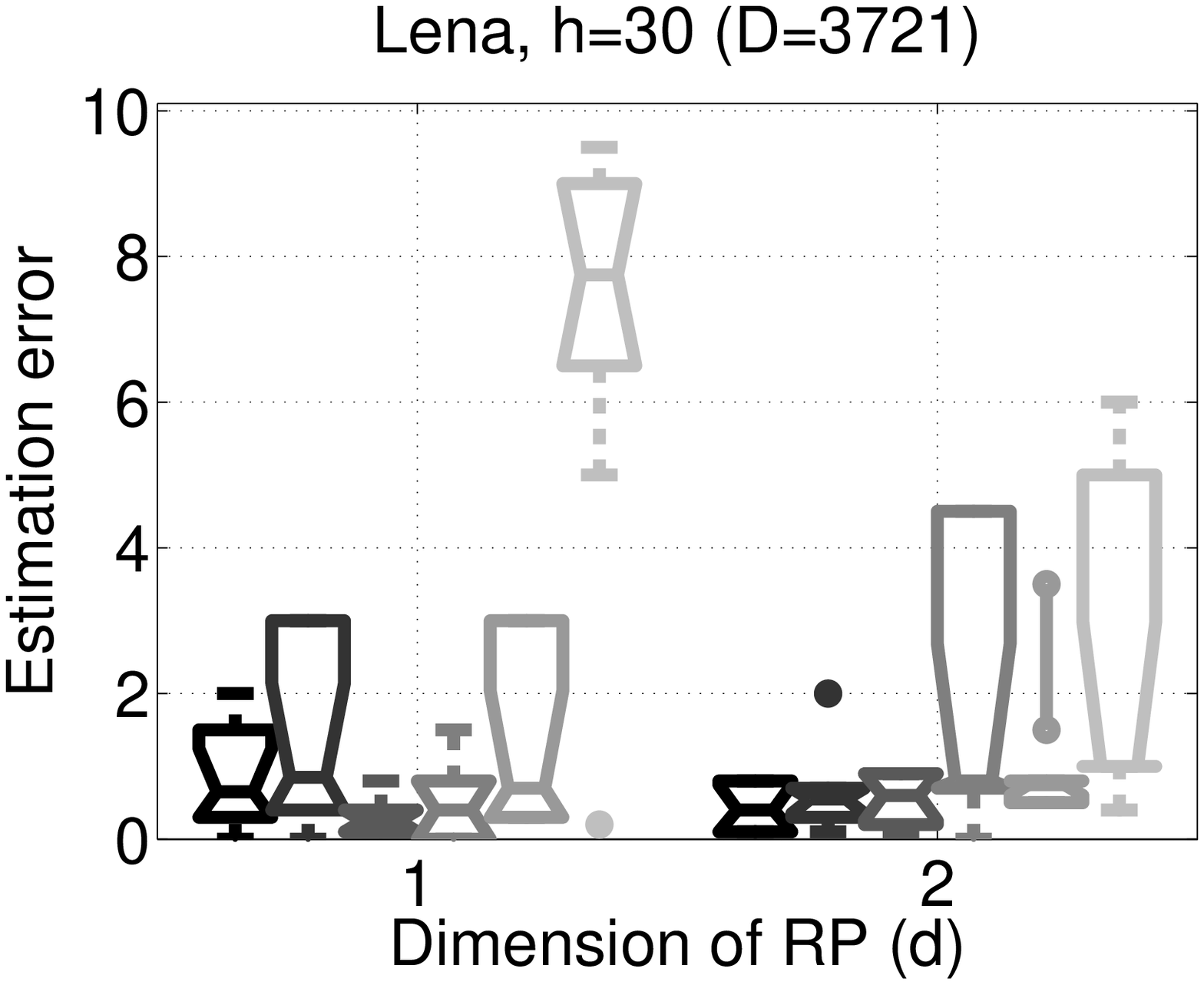}}%
  \subfloat[][]{\includegraphics[width=7.15cm]{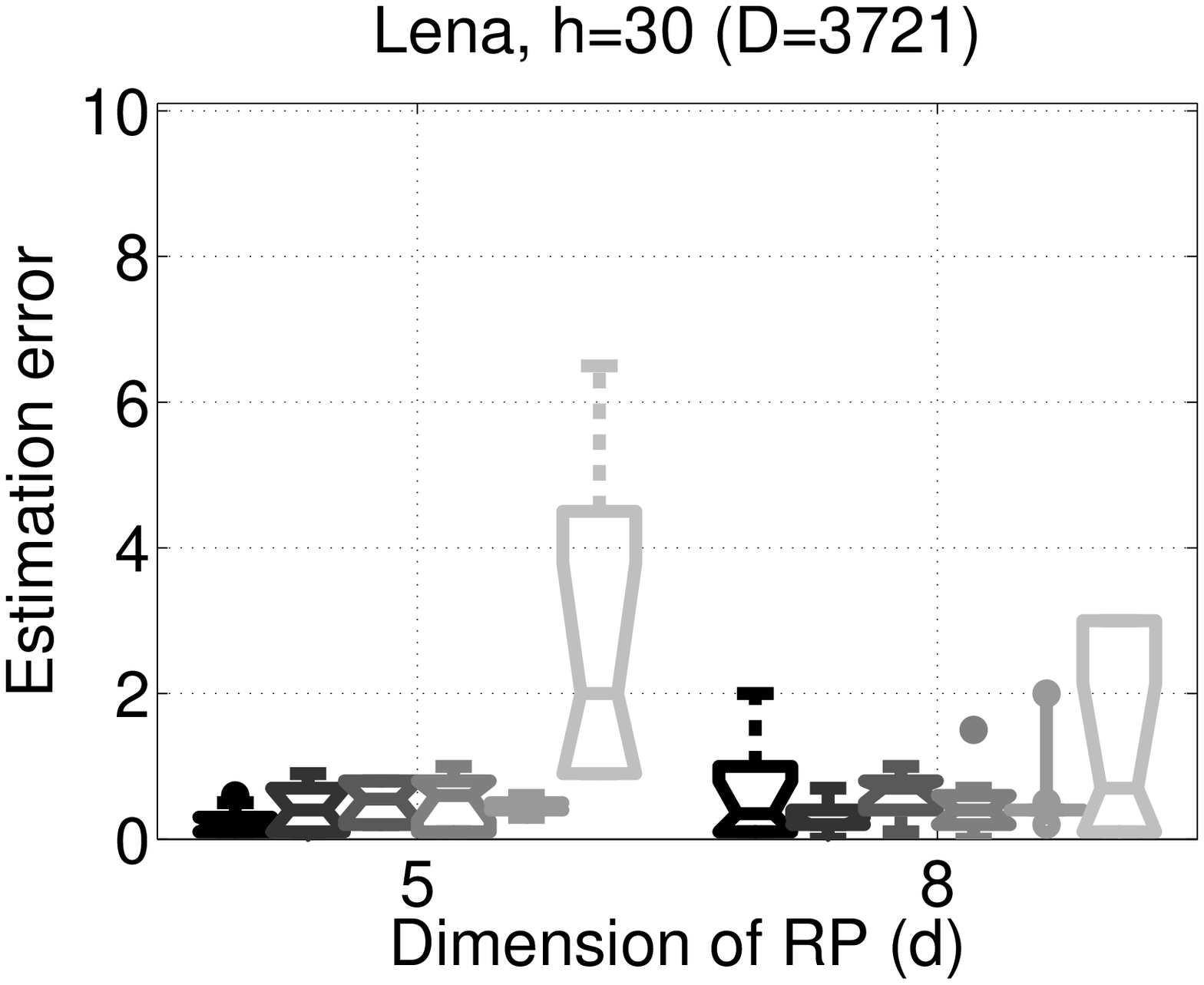}}%
  \caption[Estimation error: \emph{Lena} dataset, \emph{kdp} method.]{Estimation error as a function of the RP dimension $d$ 
   on the \emph{Lena} dataset for different $G$ group sizes. Method: \emph{kdp}. 
    (a)-(b): $h=10$. (c)-(d): $h=30$. First column: $d=1$, $2$. Second column: $d=5$, $8$.}%
  \label{fig:lena:h10,30:kdpee}%
\end{figure}

%----------------
%estimatiion error: mandrill:

\begin{figure}%[h!]
  \centering%
  \subfloat[][]{\includegraphics[width=7.15cm]{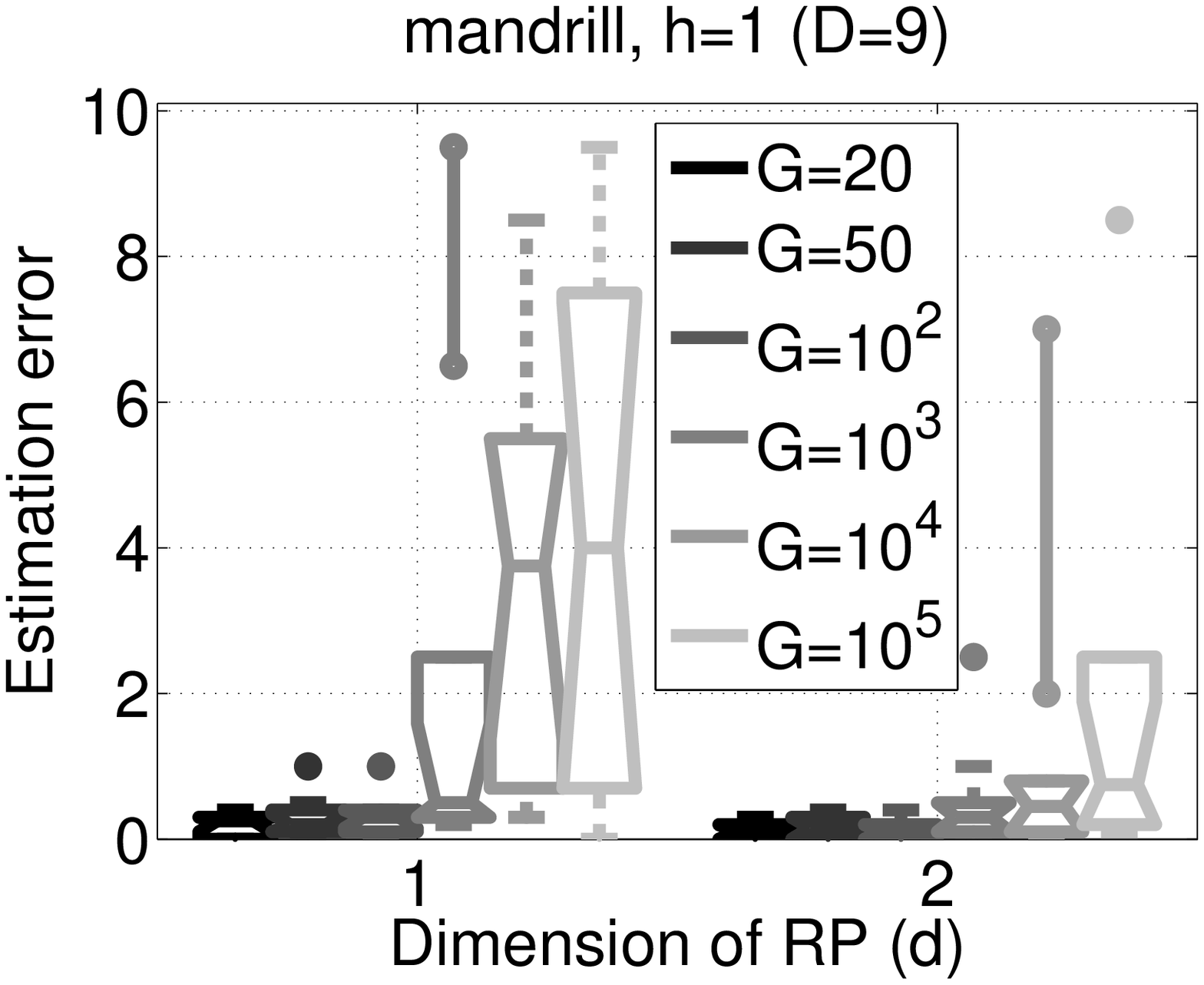}}%
  \subfloat[][]{\includegraphics[width=7.15cm]{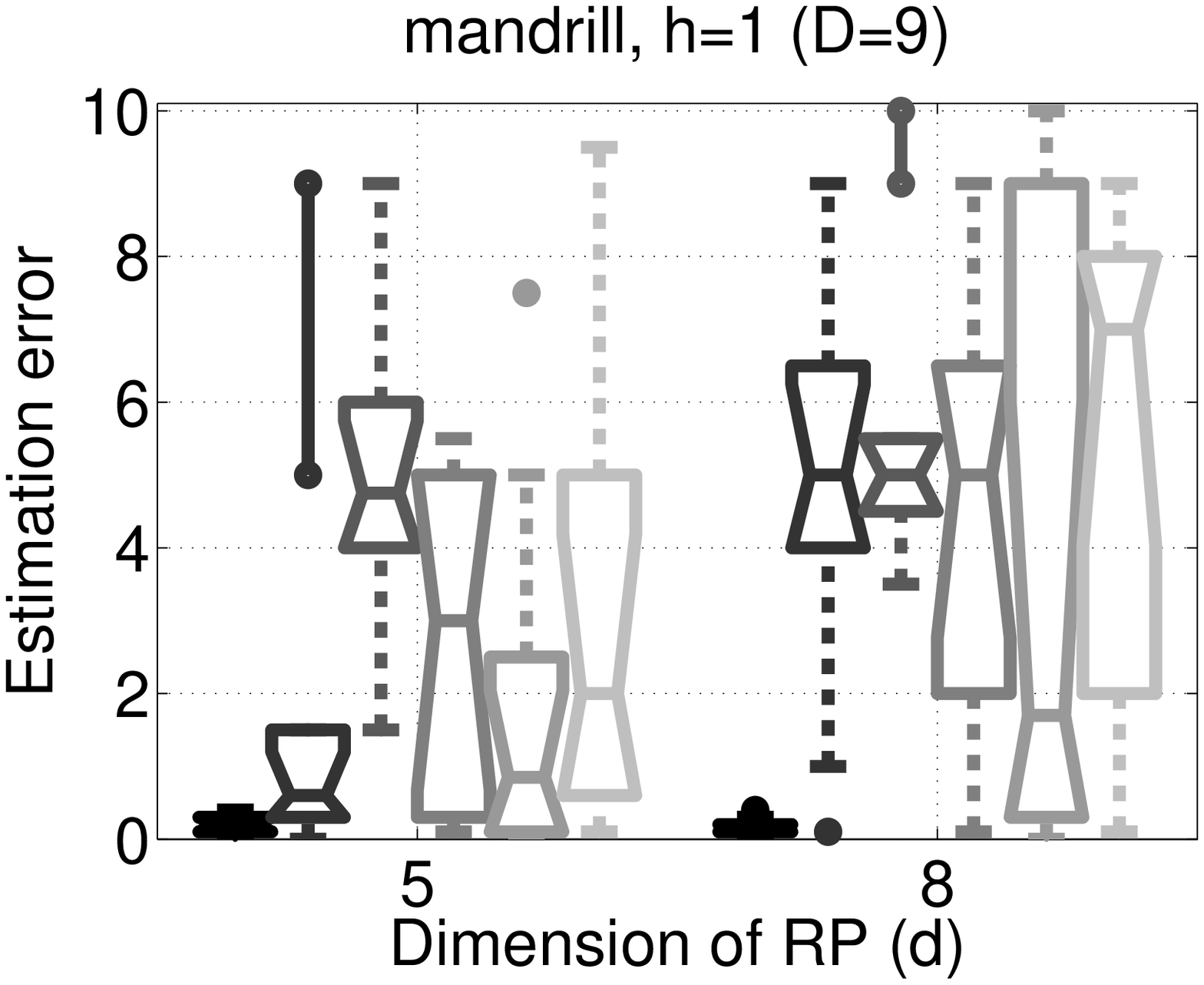}}\\%
  \subfloat[][]{\includegraphics[width=7.15cm]{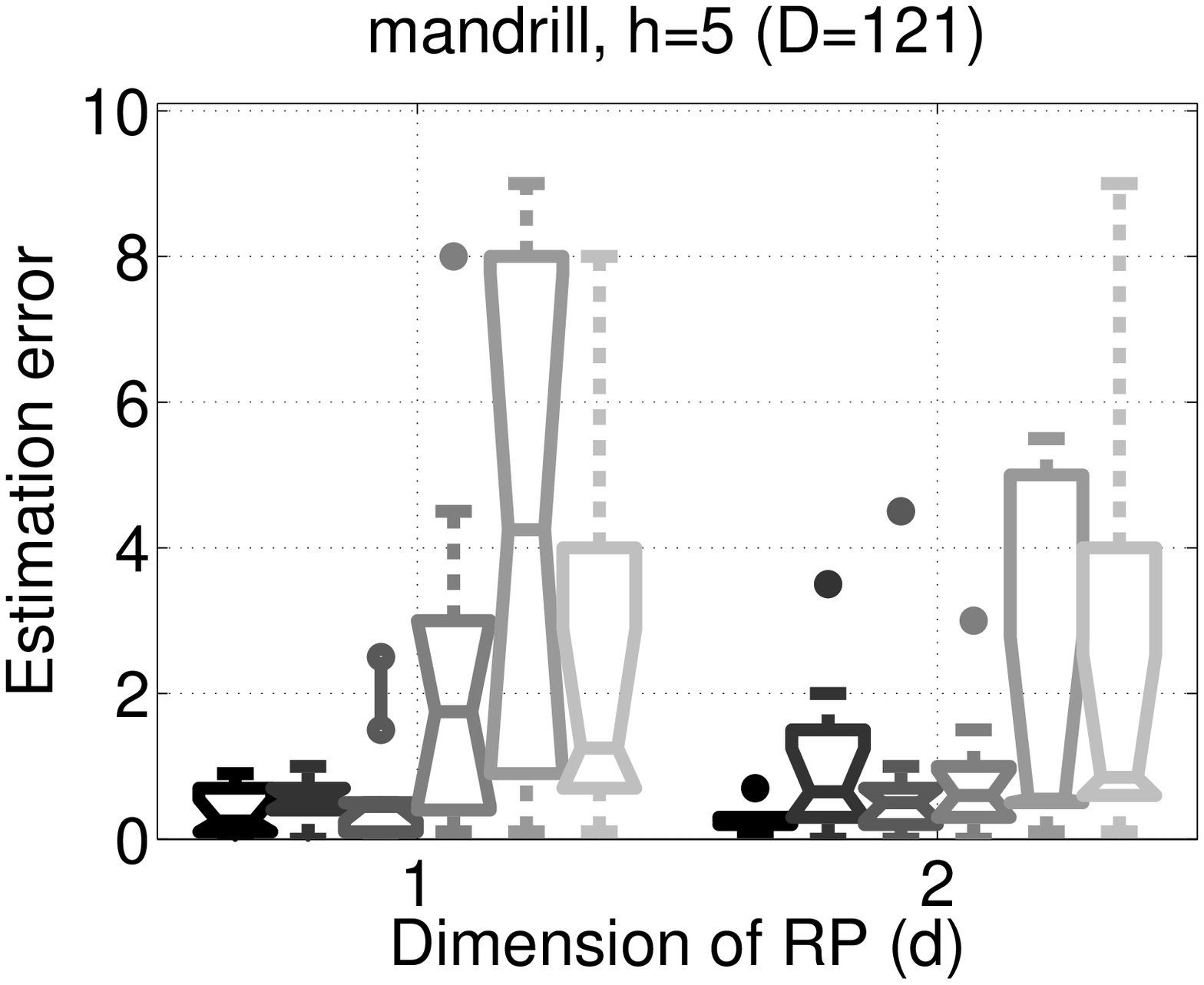}}%
  \subfloat[][]{\includegraphics[width=7.15cm]{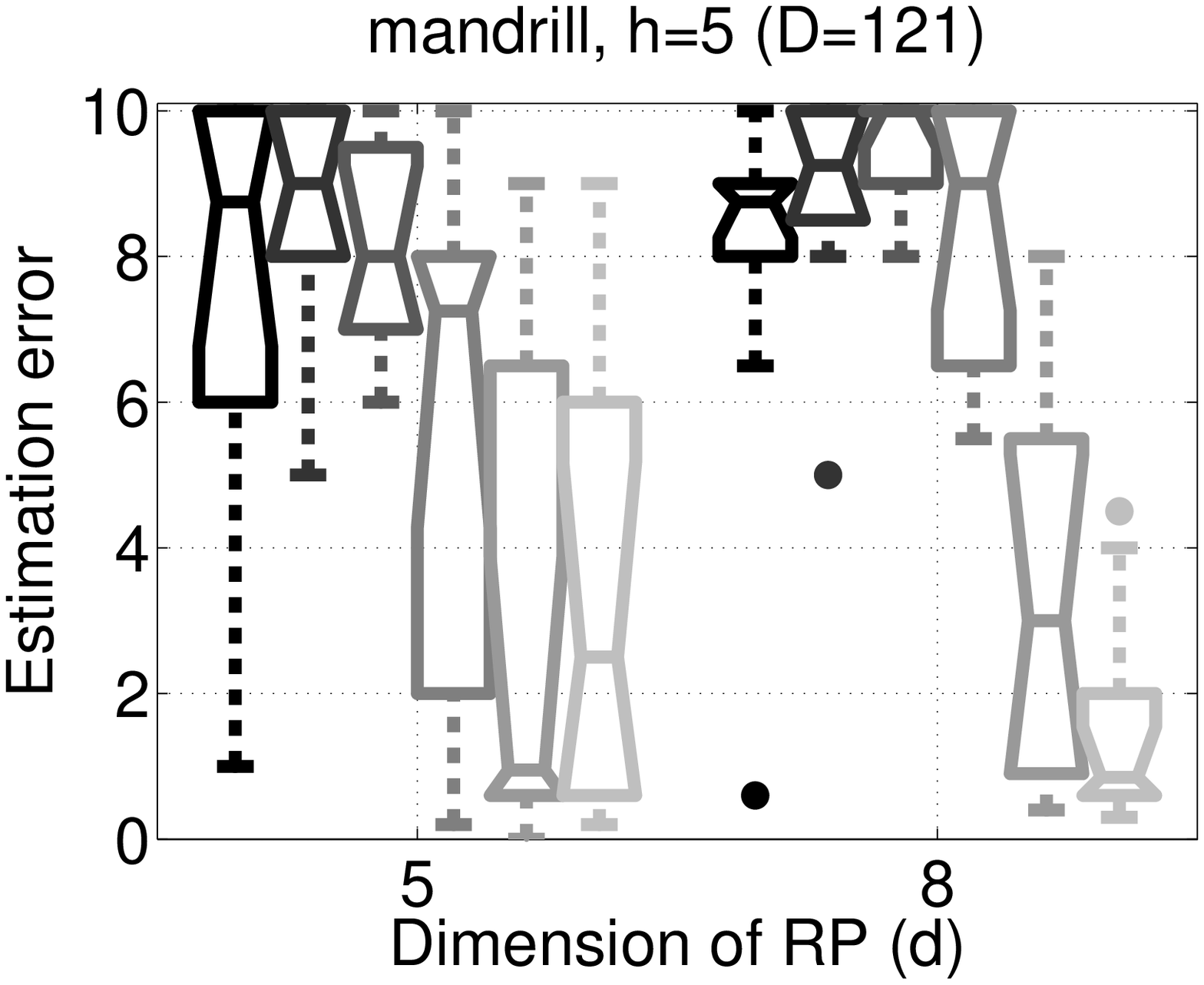}}%
  \caption[Estimation error: \emph{mandrill} dataset, \emph{kdp} method.]{Estimation error as a function of 
  the RP dimension $d$  on the \emph{mandrill} dataset for different $G$ group sizes. Method: \emph{kdp}. 
    (a)-(b): neighbor size $h=1$. (c)-(d): $h=5$. First column: $d=1$, $2$. Second column: $d=5$, $8$.}%
  \label{fig:mandrill:h1,5:kdpee}%
\end{figure}

\begin{figure}%[h!]
  \centering%
  \subfloat[][]{\includegraphics[width=7.15cm]{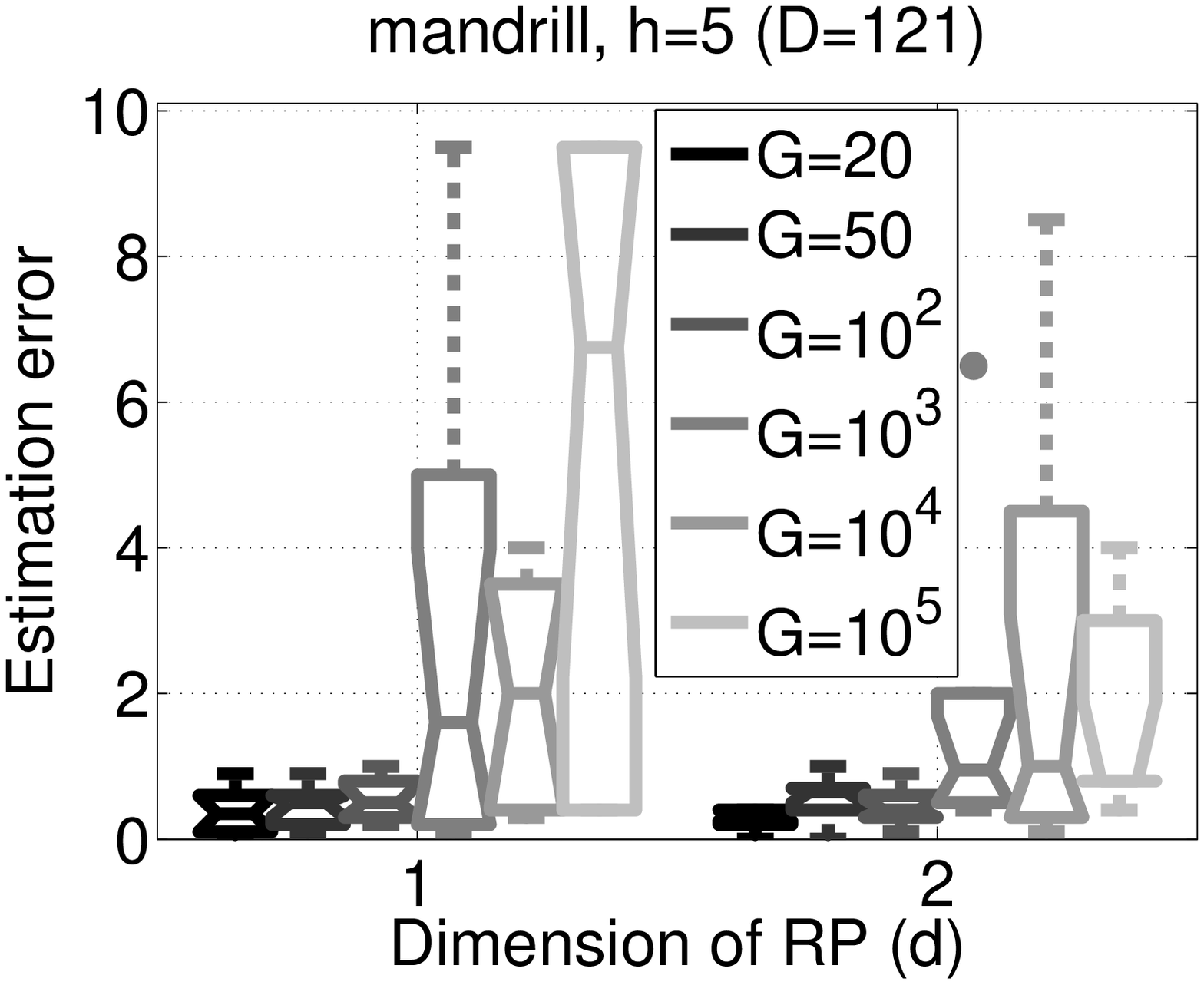}}%
  \subfloat[][]{\includegraphics[width=7.15cm]{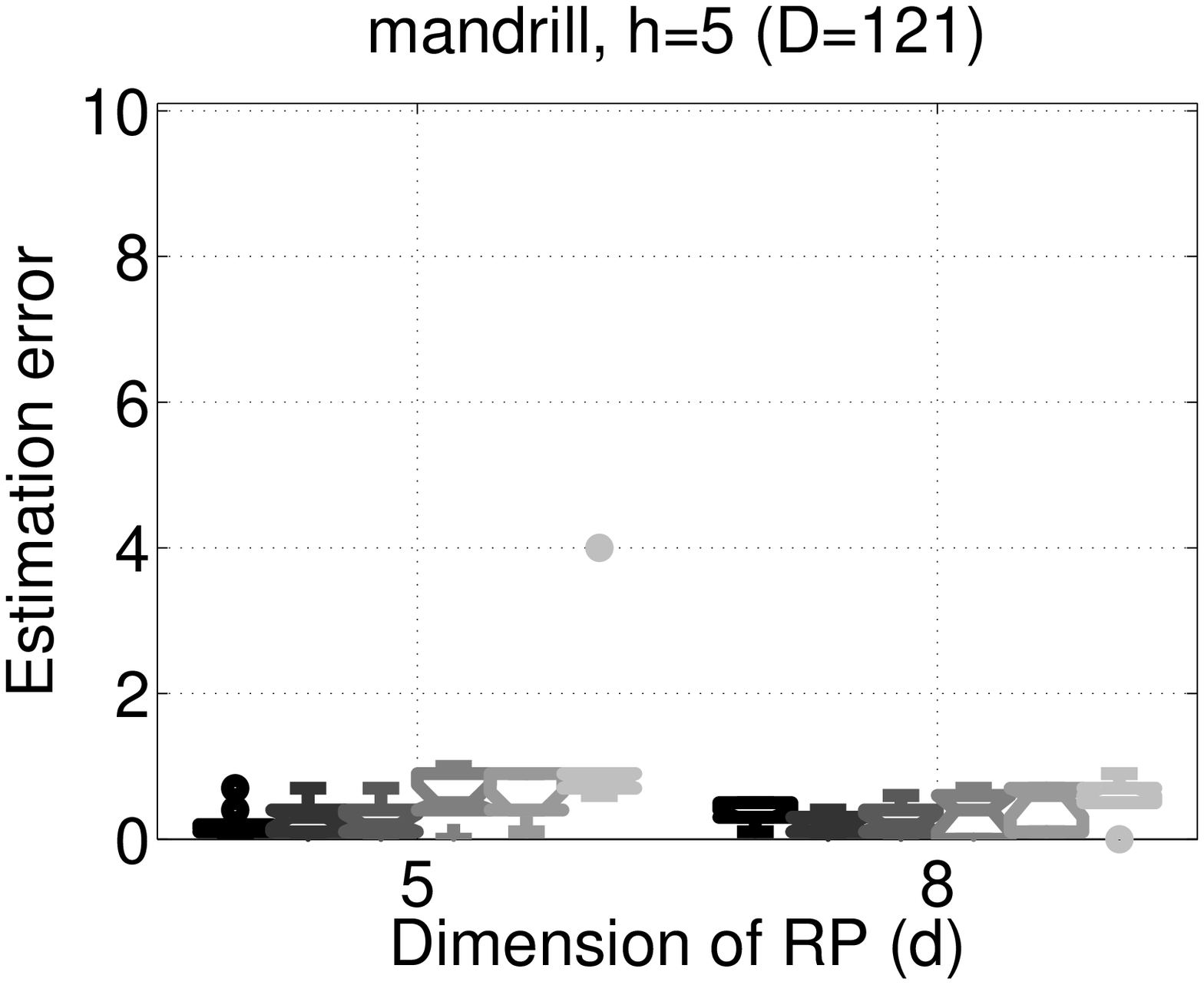}}\\%
  \subfloat[][]{\includegraphics[width=7.15cm]{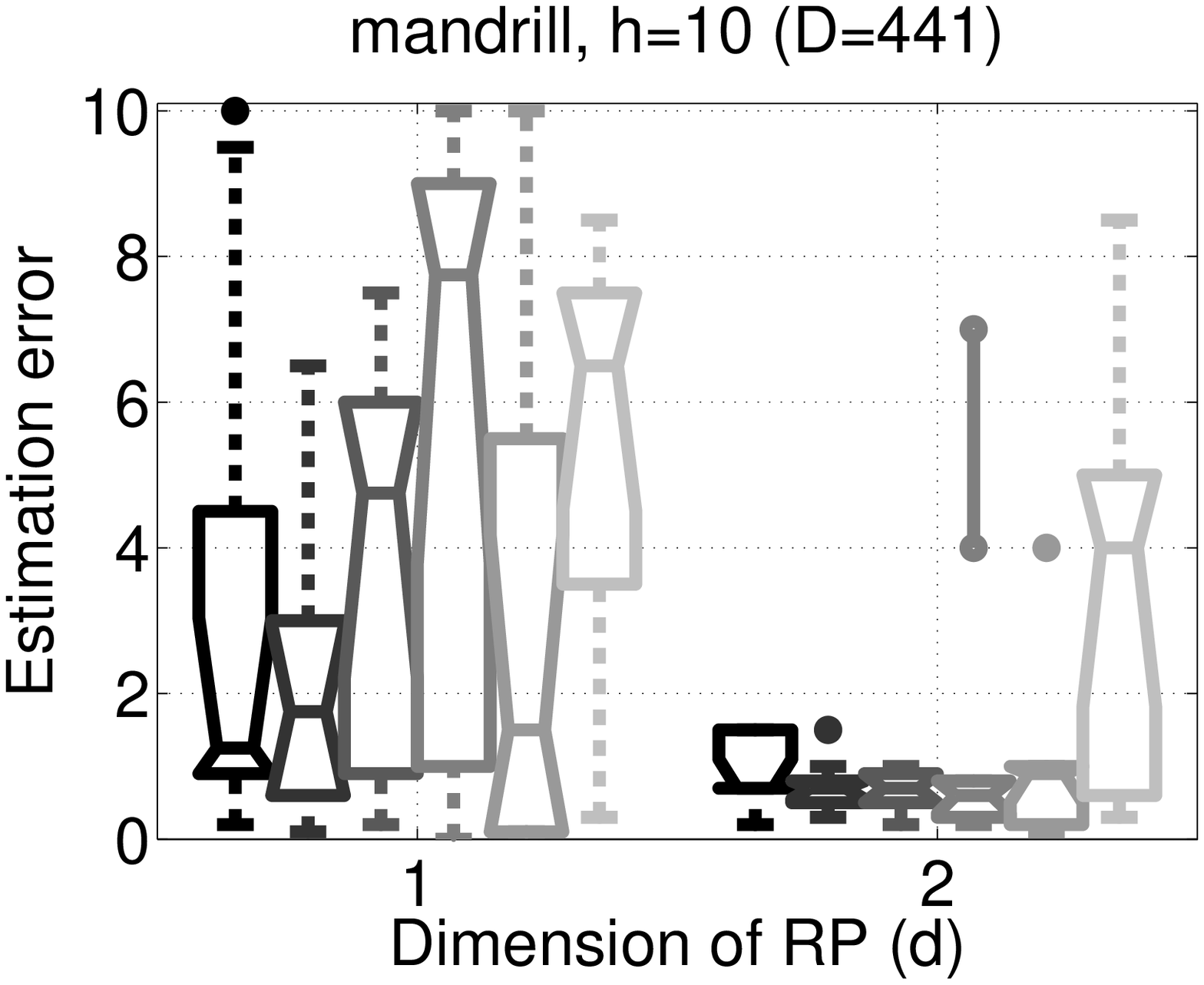}}%
  \subfloat[][]{\includegraphics[width=7.15cm]{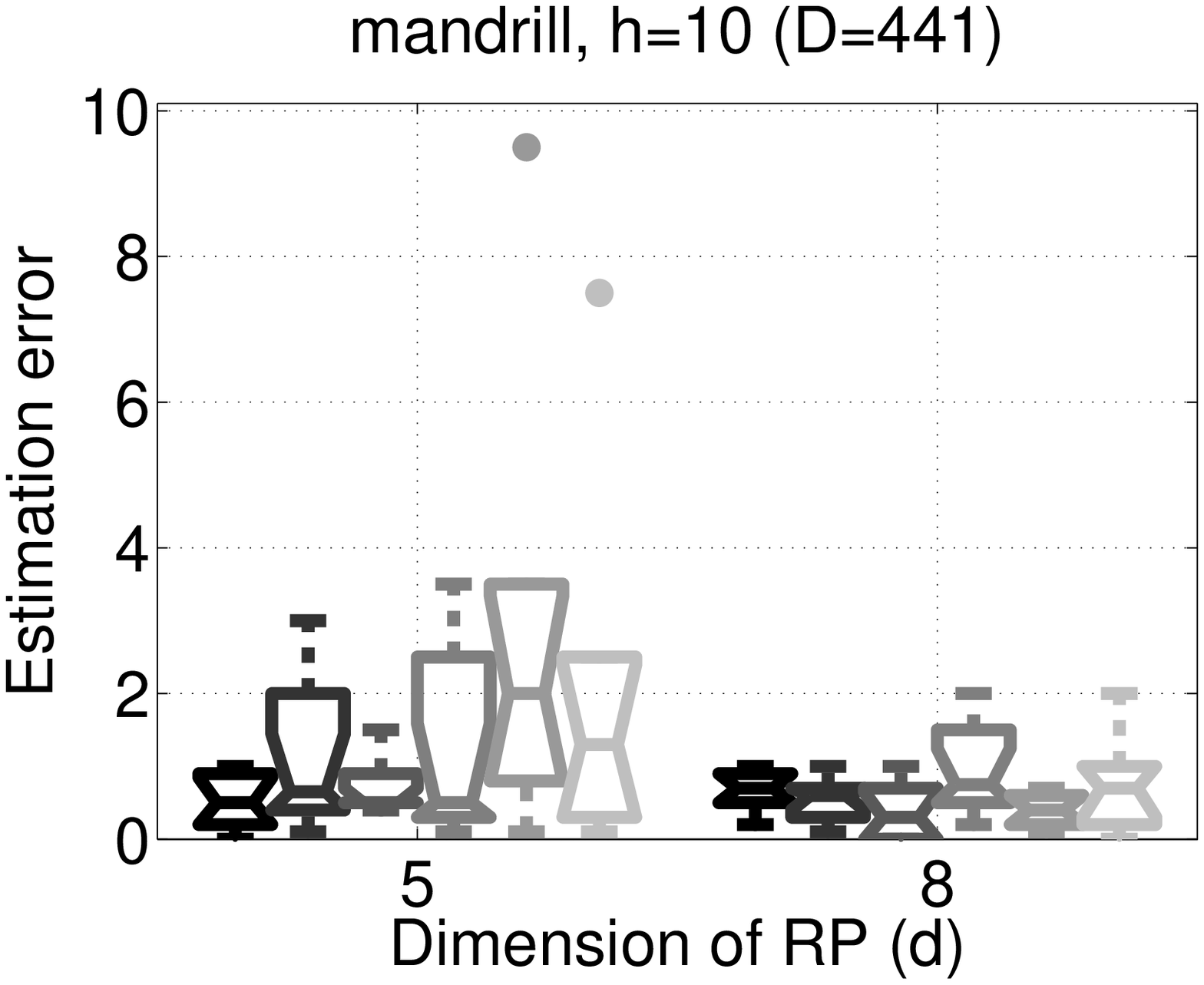}}%
  \caption[Estimation error: \emph{mandrill} dataset, \emph{kNN$_{k}$} method.]{Estimation error as a function of 
  the RP dimension $d$  on the \emph{mandrill} dataset for different $G$ group sizes. Method: \emph{kNN$_k$}. 
    (a)-(b): neighbor size $h=5$. (c)-(d): $h=10$. First column: $d=1$, $2$. Second column: $d=5$, $8$.}%
  \label{fig:mandrill:h5,10:kNNk}%
\end{figure}

\begin{figure}%[h!]
  \centering%
  \subfloat[][]{\includegraphics[width=7.15cm]{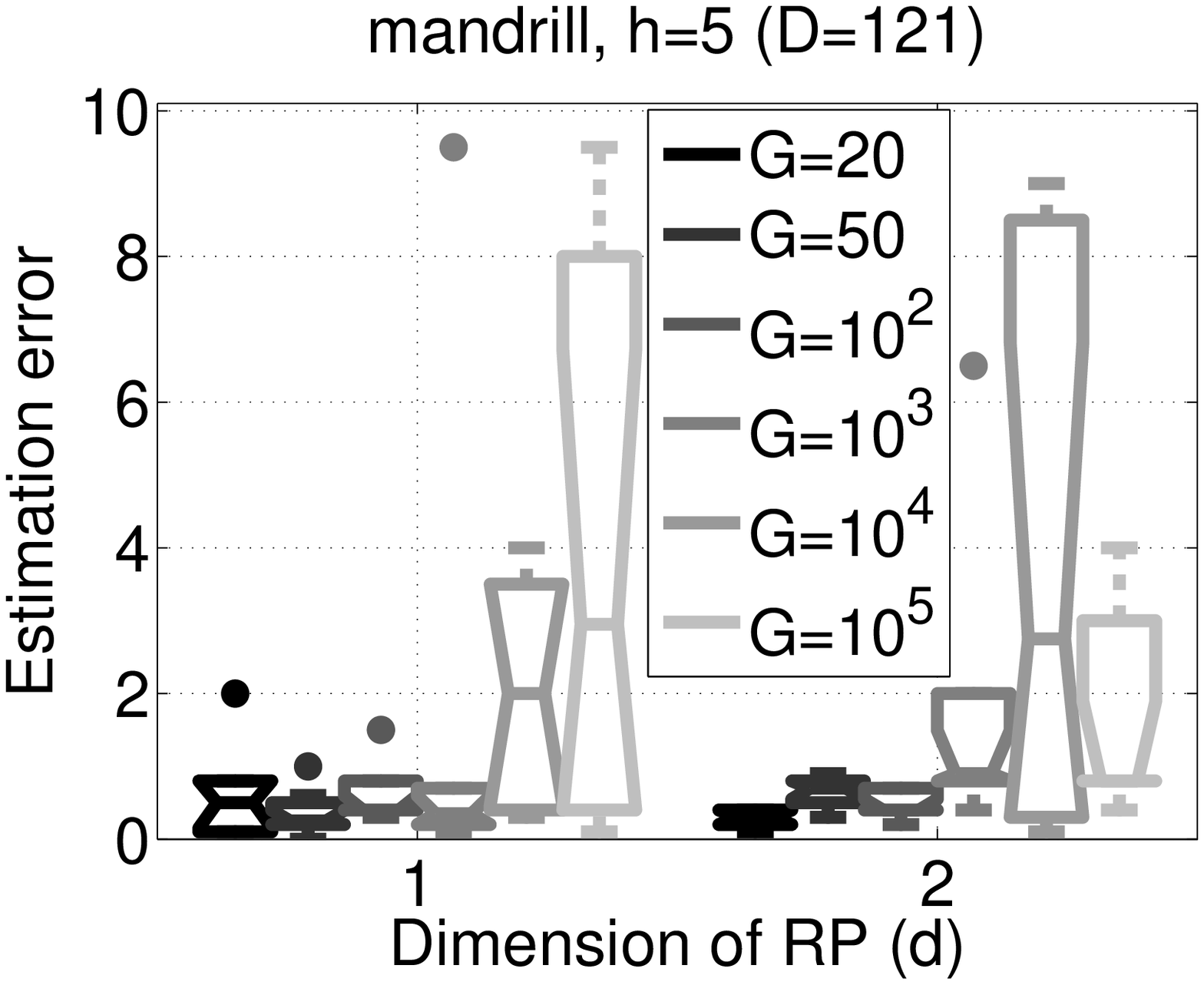}}%
  \subfloat[][]{\includegraphics[width=7.15cm]{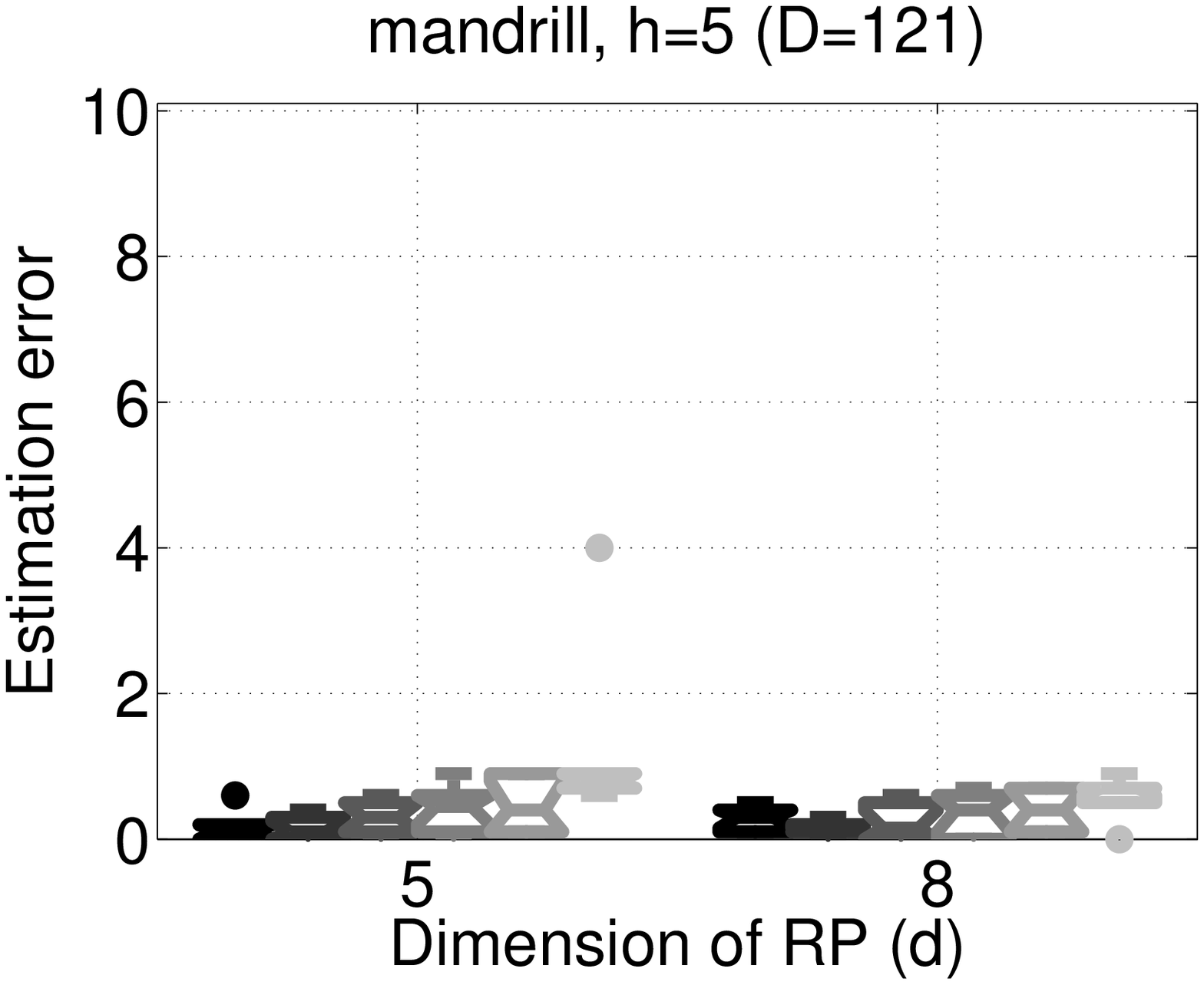}}\\%
  \subfloat[][]{\includegraphics[width=7.15cm]{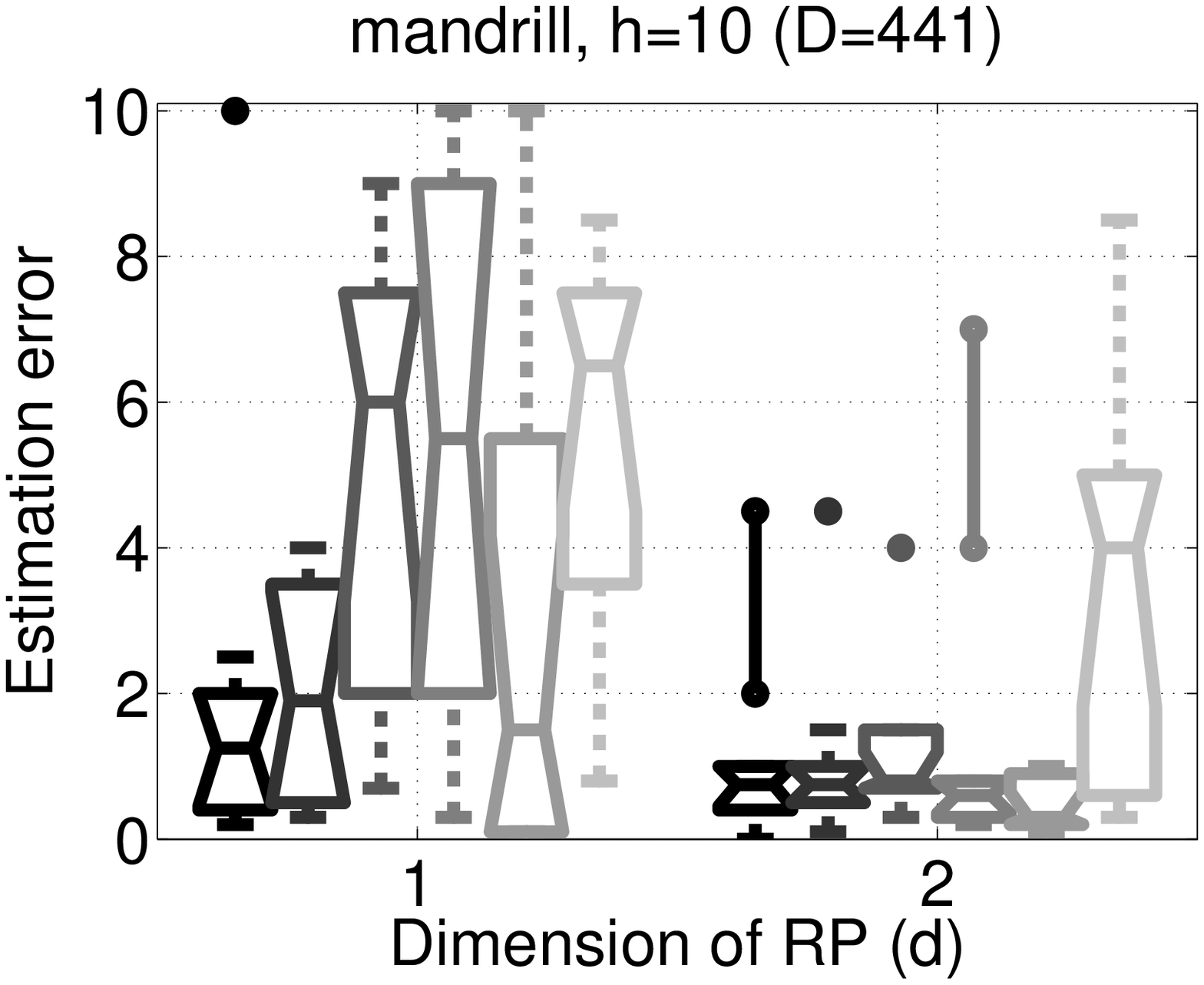}}%
  \subfloat[][]{\includegraphics[width=7.15cm]{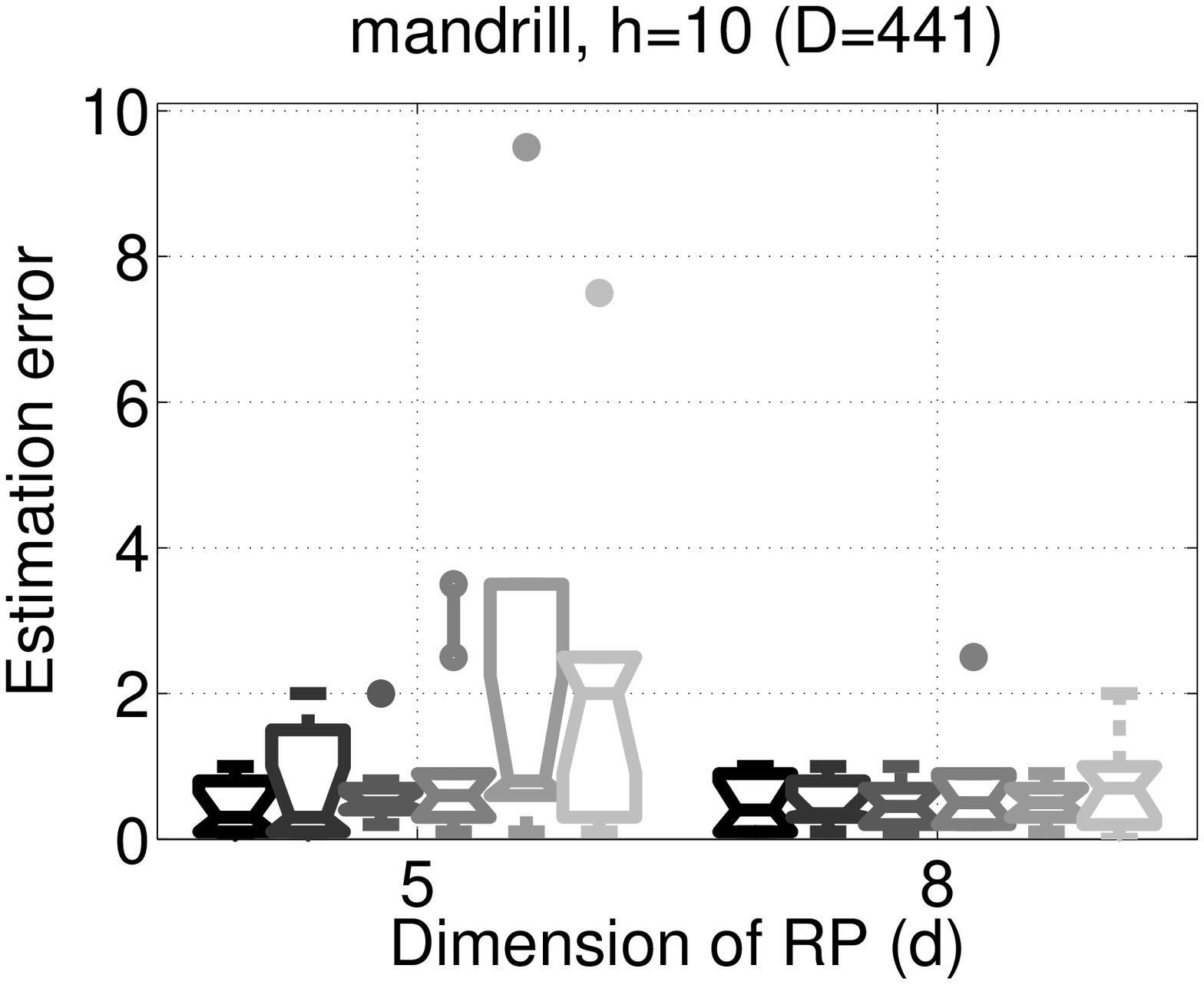}}%
  \caption[Estimation error: \emph{mandrill} dataset, \emph{kNN$_{1-k}$} method.]{Estimation error as a function of 
  the RP dimension $d$  on the \emph{mandrill} dataset for different $G$ group sizes. Method: \emph{kNN$_{1-k}$}. 
    (a)-(b): neighbor size $h=5$. (c)-(d): $h=10$. First column: $d=1$, $2$. Second column: $d=5$, $8$.}%
  \label{fig:mandrill:h5,10:kNN1tok}%
\end{figure}

\begin{figure}%[h!]
  \centering%
  \subfloat[][]{\includegraphics[width=7.15cm]{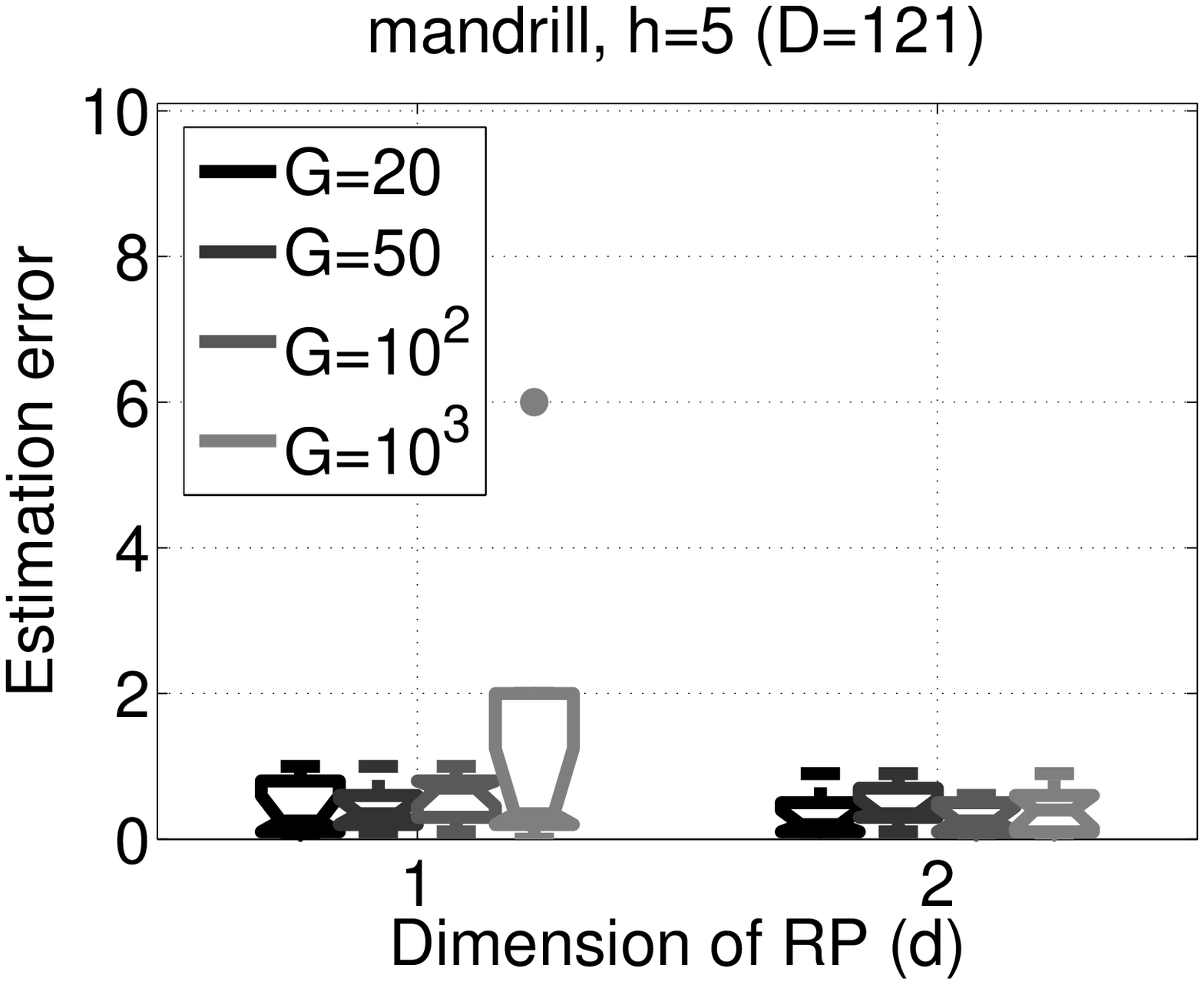}}%
  \subfloat[][]{\includegraphics[width=7.15cm]{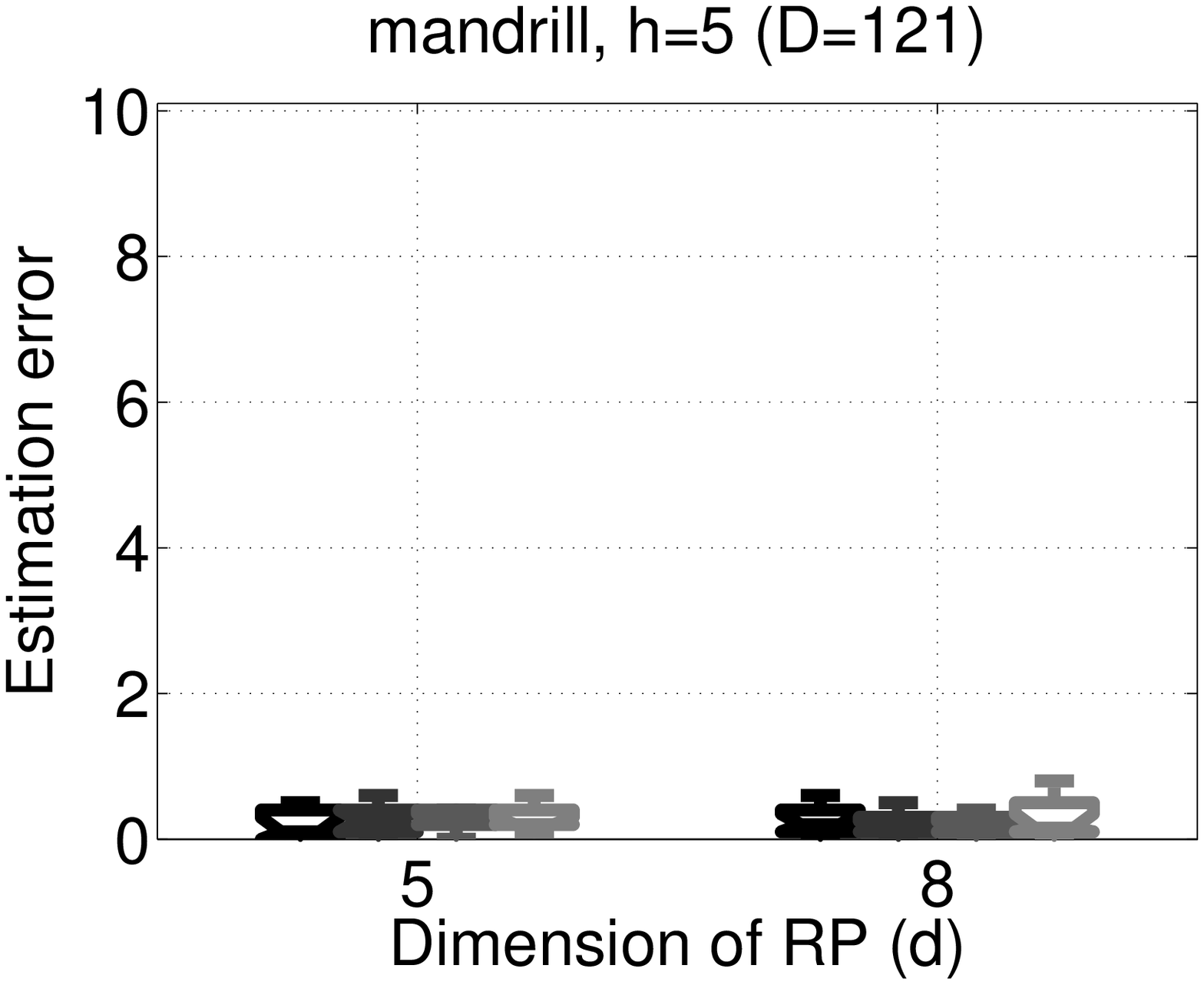}}\\%
  \subfloat[][]{\includegraphics[width=7.15cm]{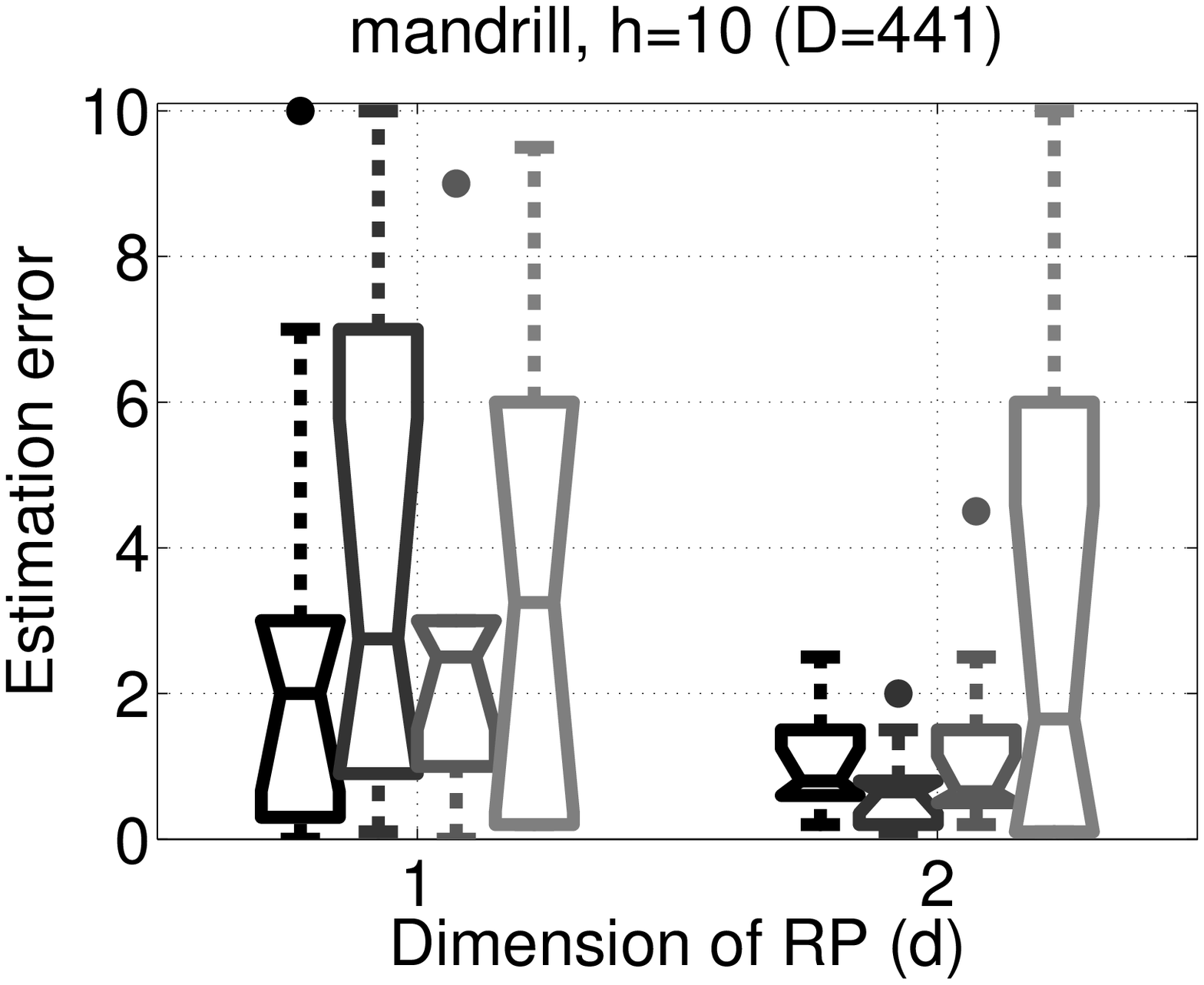}}%
  \subfloat[][]{\includegraphics[width=7.15cm]{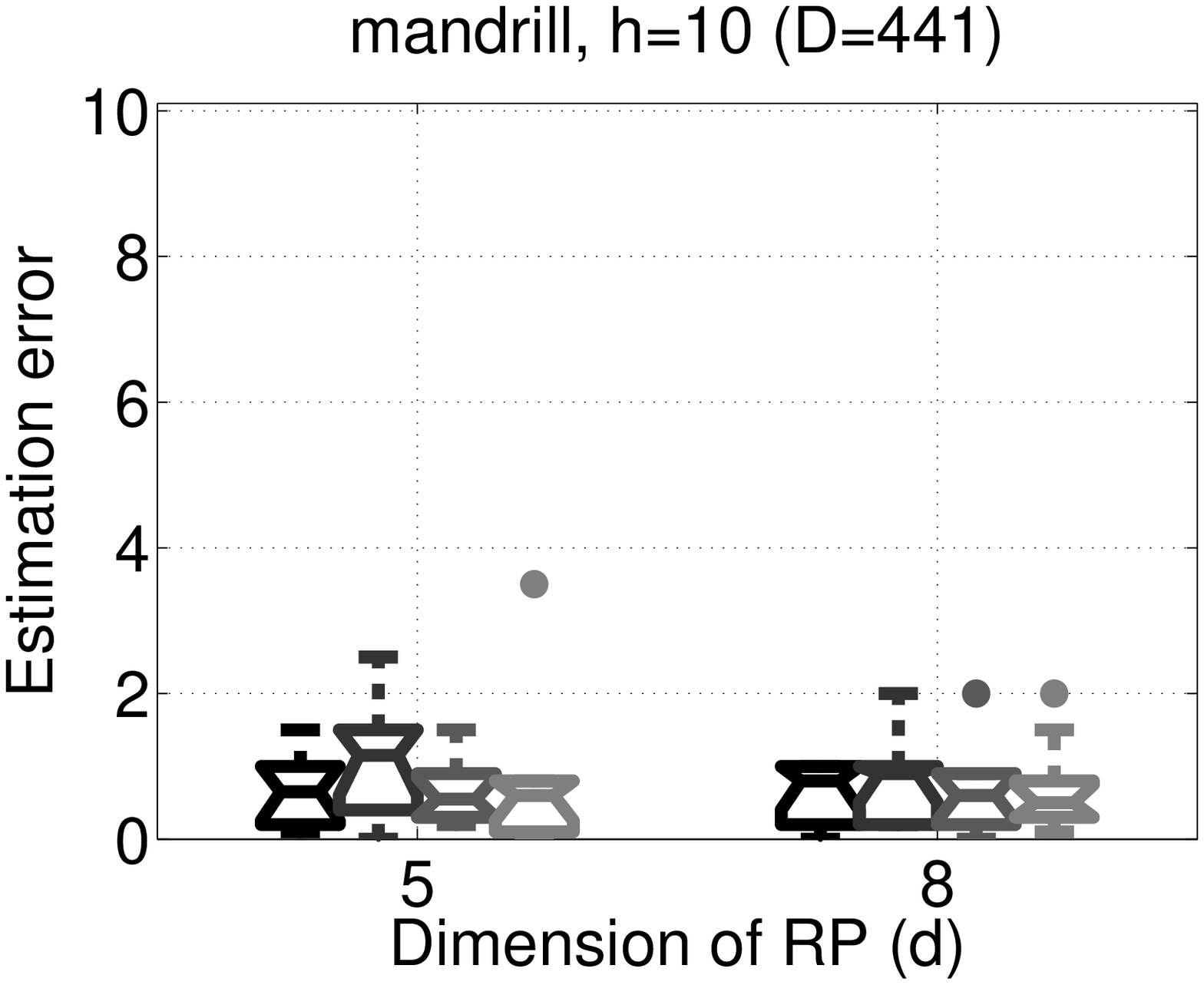}}%
  \caption[Estimation error: \emph{mandrill} dataset, \emph{MST} method.]{Estimation error as a function of 
   the RP dimension $d$  on the \emph{mandrill} dataset for different $G$ group sizes. Method: \emph{MST}. 
   (a)-(b): neighbor size $h=5$. (c)-(d): $h=10$. First column: $d=1$, $2$. Second column: $d=5$, $8$.}% %magic_const=2
  \label{fig:mandrill:h5,10:MST}%
\end{figure}

\begin{figure}%[h!]
  \centering%
  \subfloat[][]{\includegraphics[width=7.15cm]{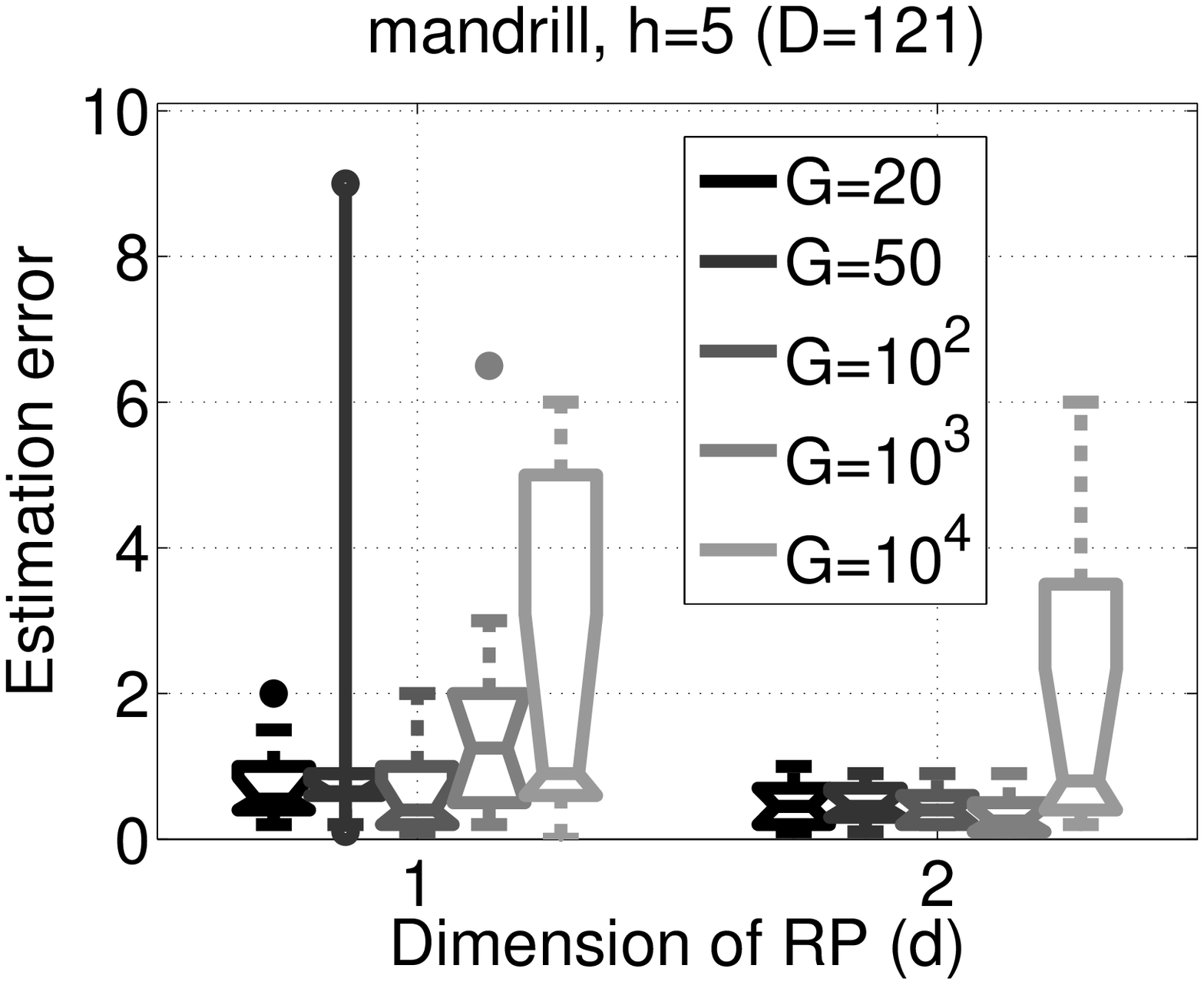}}%
  \subfloat[][]{\includegraphics[width=7.15cm]{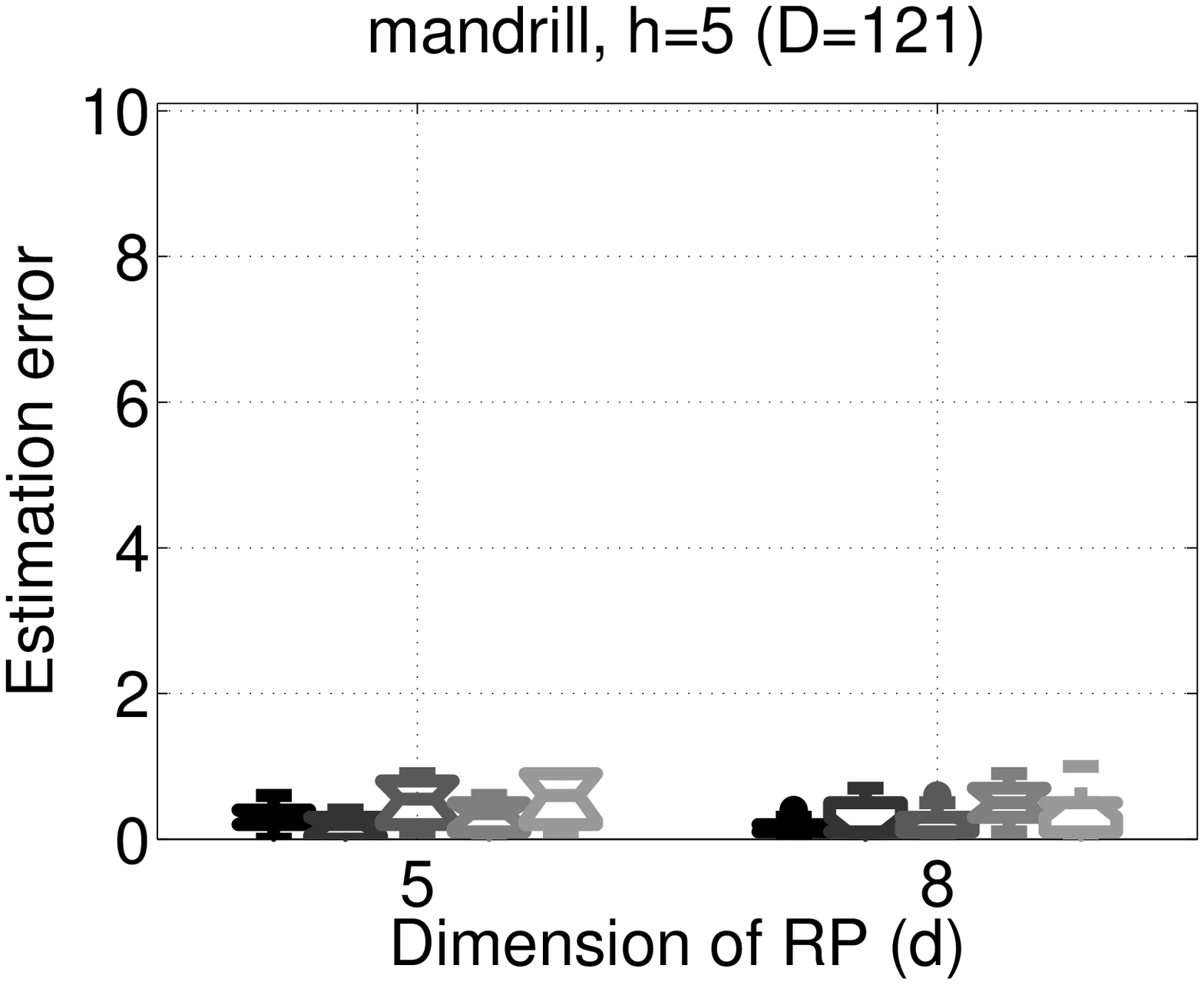}}\\%
  \subfloat[][]{\includegraphics[width=7.15cm]{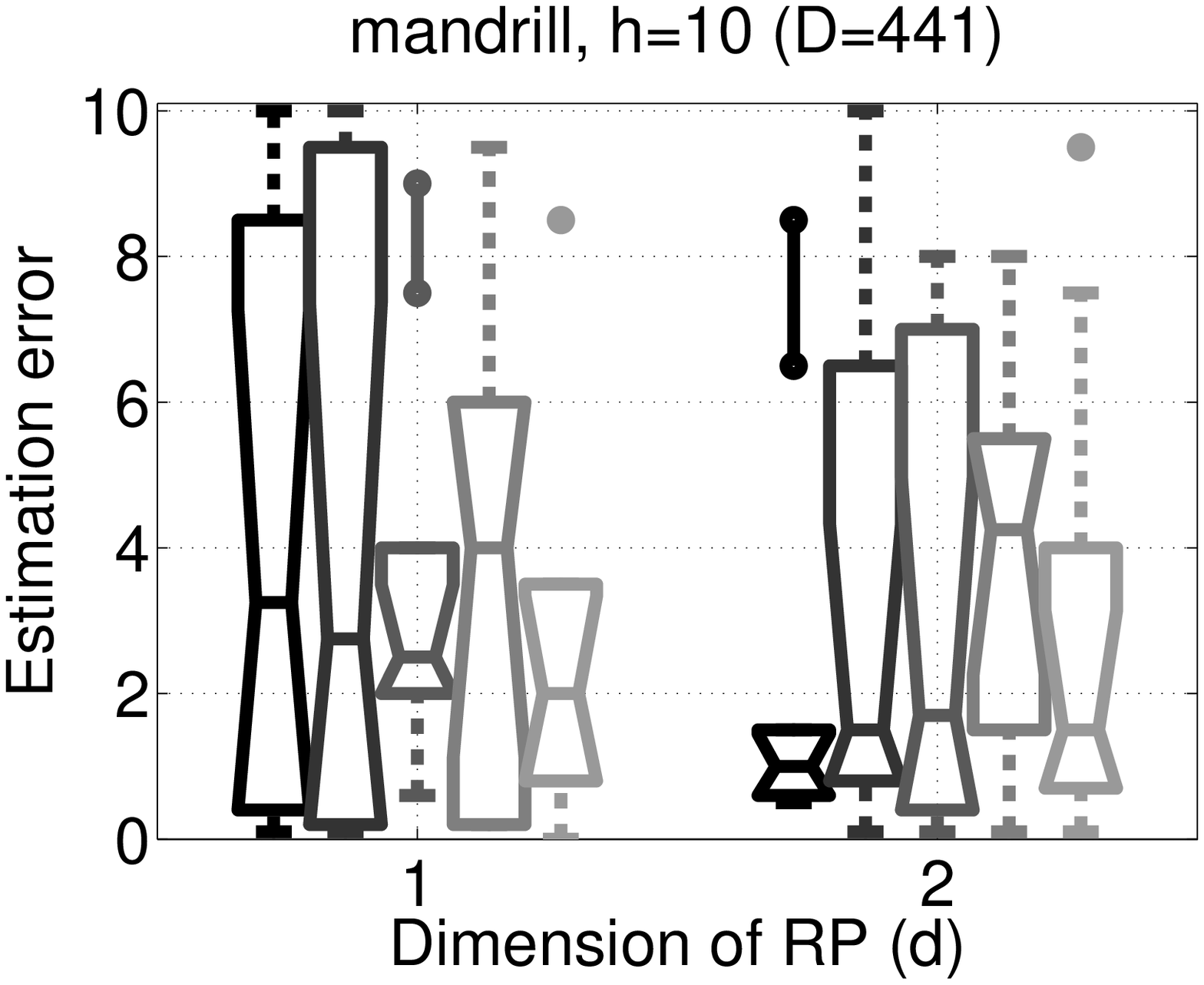}}%
  \subfloat[][]{\includegraphics[width=7.15cm]{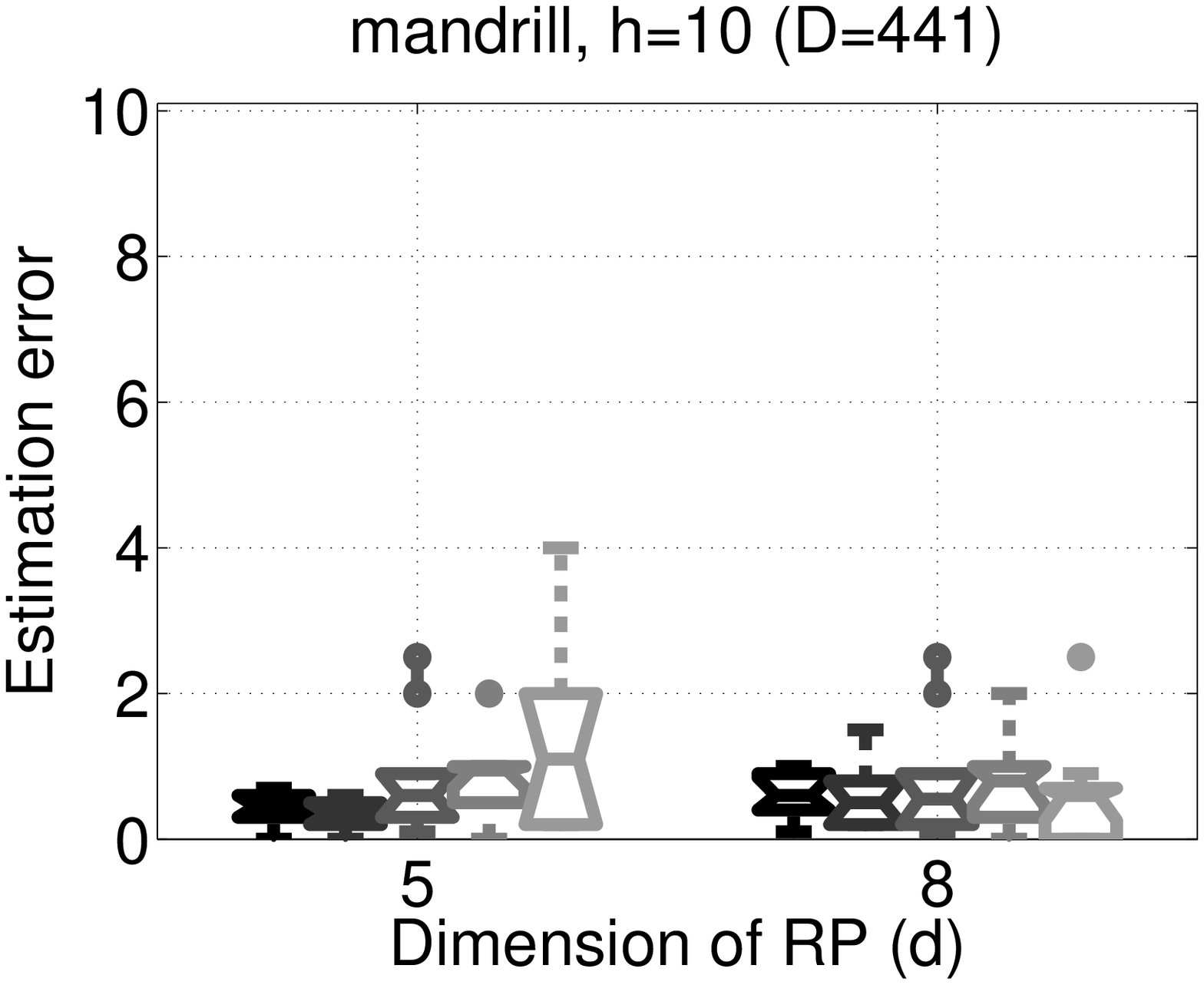}}%
  \caption[Estimation error: \emph{mandrill} dataset, \emph{wkNN} method.]{Estimation error as a function of 
   the RP dimension $d$  on the \emph{mandrill} dataset for different $G$ group sizes. Method: \emph{wkNN}. 
   (a)-(b): neighbor size $h=5$. (c)-(d): $h=10$. First column: $d=1$, $2$. Second column: $d=5$, $8$.}%
  \label{fig:mandrill:h5,10:weightedkNN}%
\end{figure}

%-----------------

\begin{table}%[h!]
    \centering
    \begin{tabular}{|r||r|r|r|r|}
    \hline
			     & $G=20$ & $G=50$ & $G=100$ & $G=1000$\\
      \hline\hline
      kdp                  & $1.01$ & $1.27$ & $1.29$  & $2.92$\\\hline
      kNN$_{1-k}$ (kNN$_{k}$)& $2.25$ & $5.71$ & $12.65$ & $125.84$\\\hline
      MST                    & $1.15$ & $1.36$ & $1.65$  & $1.82$\\\hline
      wkNN            & $1.30$ & $2.59$ & $4.88$   & $16.05$\\
    \hline
    \end{tabular}
    \caption[Computation time versus raw data based methods.]{Computation time versus raw data based method. 
    Value $z>1$, means $z$ times improvement in computation time over the method not applying dimension reduction. Dataset: \emph{mandrill}. Baseline: RP dimension $d=2$. Neighbor size: $h=5$.}
    \label{tab:elapsed-vs-d2}
\end{table}

\end{document}